\documentclass[aps,superscriptaddress,twocolumn]{revtex4-2}
\usepackage[utf8]{inputenc}
\usepackage{amsthm,amssymb,amsmath}
\usepackage{graphicx,xcolor}
\usepackage[colorlinks=true,linkcolor=blue,citecolor=blue,urlcolor=blue,linktocpage=true]{hyperref}

\usepackage{microtype}
\usepackage{booktabs}
\usepackage{float}
\usepackage{multirow}
 \usepackage{siunitx}
\definecolor{cg}{rgb}{0,0.6,0}

\begin{document}
\title{Exact description of fermionic reservoirs via purified damped ancillary fermions}

\author{Pengfei Liang}
\email{pfliang@imu.edu.cn}
\affiliation{Research Center for Quantum Physics and Technologies, Inner Mongolia University, Hohhot 010021, China}
\affiliation{School of Physical Science and Technology, Inner Mongolia University, Hohhot 010021, China}

\author{Neill Lambert}
\email{nwlambert@gmail.com}
\affiliation{RIKEN Center for Quantum Computing (RQC), Wakoshi, Saitama 351-0198, Japan}

\author{Mauro Cirio}
\email{cirio.mauro@gmail.com}
\affiliation{Graduate School of China Academy of Engineering Physics, Haidian District, Beijing, 100193, China}

\date{\today}
\begin{abstract}
We present a method for the modeling of fermionic reservoirs using a new class of ancillary damped fermions, dubbed purified pseudofermions, which exhibit unusual free correlations. We show that this key feature, when combined with existing efficient decomposition algorithms for the reservoir correlation functions, enables the development of an easily implementable and accurate scheme for constructing effective models of fermionic reservoirs. We numerically demonstrate the validity, accuracy, efficiency and potential use of our method by studying the particle transport of spinless fermions in a one-dimensional chain. Beyond its utility as a quantum impurity solver, our method holds promise for addressing a wide range of problems involving extended systems in fields like  quantum transport, quantum thermodynamics, thermal engines and nonequilibrium phase transitions.  
\end{abstract}

\maketitle

\section{Introduction}\label{sec:intro}

In solid-state physics or, more generally, condensed matter physics, physical systems consisting of a continuum of electronic states, such as metals, are often theoretically referred to as fermionic reservoirs~\cite{Datta_1995}. They appear ubiquitously in a range of theoretical problems, from electronic transport through single-molecule junctions~\cite{10.1063/1.5003306,Gehring2019,RevModPhys.92.035001} or quantum dots~\cite{RevModPhys.74.1283,RevModPhys.75.1}, to the Kondo resonance related to magnetic impurities in conducting metals~\cite{RevModPhys.55.331,Hewson_1993} or quantum thermodynamics~\cite{PhysRevE.76.031105,Topp_2015,Josefsson2018,Mosso2019,PhysRevX.10.031040}. Considerable efforts have been devoted to the understanding of transport or many-body properties in such fermionic open quantum systems, where a small system consisting of a few electronic levels couples to its surrounding electronic reservoirs, which in turn have led to significant advances in molecular electronics and nanoscale functional devices. 

In order to properly address the strong hybridization between the few-level system and the reservoirs it couples to, methods that can go beyond the conventional weak-coupling limit and Born-Markov approximation underlying the regular quantum master equation approach~\cite{PhysRevLett.97.166801,PhysRevB.77.195416,PhysRevB.79.205303,PhysRevB.83.115414} are needed. Several existing methods, such as the numerical renormalization group~\cite{PhysRevLett.95.196801,PhysRevB.74.245113}, the multilayer multiconfiguration time-dependent Hartree~\cite{10.1063/1.3173823,10.1063/1.3660206,WANG201813}, and the density-matrix renormalization group (DMRG)~\cite{PhysRevB.79.235336,PhysRevB.92.155126,PhysRevB.104.014303}, are based on the discretization of the reservoir continuum and may require special treatments to reach long-time dynamics. The hierarchical equations of motion (HEOM)~\cite{doi:10.1143/JPSJ.58.101,10.1063/1.2713104,10.1063/1.2938087,PhysRevLett.109.266403,PhysRevLett.111.086601,PhysRevB.88.235426,PhysRevB.92.085430,PhysRevB.107.195429}, continuous-time quantum Monte Carlo~\cite{PhysRevLett.100.176403,PhysRevB.79.035320,PhysRevB.79.153302,RevModPhys.83.349,PhysRevLett.116.036801,PhysRevB.100.201104}, Inchworm Monte Carlo~\cite{PhysRevLett.115.266802,PhysRevB.95.085144}, the path integral~\cite{PhysRevB.77.195316,PhysRevB.82.205323,https://doi.org/10.1002/pssb.201349187}, and several methods related to ancillary damped fermions~\cite{PhysRevLett.110.086403,PhysRevB.89.165105,PhysRevB.92.125145,PhysRevB.94.155142,Dorda_2017,PhysRevLett.121.137702,Chen_2019,PhysRevX.10.031040,PhysRevResearch.5.033011}, make use of the Gaussianity of noninteracting thermal reservoirs and have found successful applications for charge transport problems in molecular junctions~\cite{PhysRevB.88.235426,PhysRevLett.111.086601,PhysRevB.94.201407,10.1063/1.4939843,10.1063/1.4964675}. 

\begin{figure}
\includegraphics[clip,width=8.5cm]{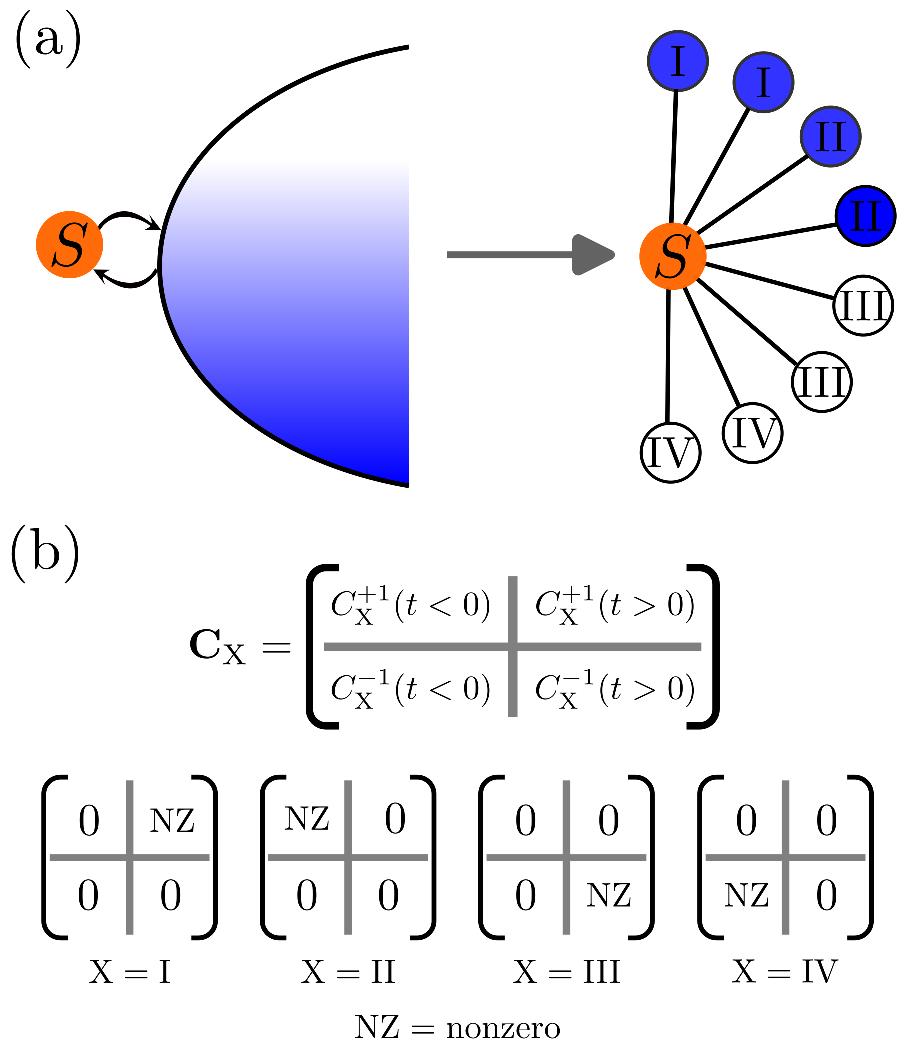}
\caption{ (a) Sketch of a fermionic open quantum system and its modeling in terms of purified pseudofermions. Darkness of the blue color corresponds to the occupation of reservoir fermionic states at the initial time. As listed in Table.~\ref{tab:ppf}, the type I and II (III and IV) purified pseudofermions are initially occupied (empty).
(b) The $2\times2$ correlation matrix $\mathbf{C}_\text{X}$ with the free correlations $C_\text{X}^{\pm1}(t)$ ($\text{X}\in\{\text{I},\text{II},\text{III},\text{IV}\}$) as entries. In this representation, $\mathbf{C}_\text{X}$ has only one nonvanishing entry for each type of purified pseudofermions. 
}\label{fig:schematics}
\end{figure}

In this work, we present a novel method based on the pseudofermion approach~\cite{PhysRevResearch.5.033011}. Specifically, inspired by our previous work~\cite{liang2024purifiedinputoutputpseudomodemodel,c91x-bhqw}, we introduce a new class of damped ancillary fermions, which we call {\it purified pseudofermions}. As sketched in Fig.~\ref{fig:schematics}(a), the environmental effects of the original fermionic reservoir can be modeled by an effective one made of such purified pseudofermions, which can be further classified into four types, labeled as type I, II, III and IV, respectively. More concretely, the classification is based on the structure of their $2\times2$ free correlation matrix ($\text{X}\in\{\text{I},\text{II},\text{III},\text{IV}\}$)
\begin{equation}\label{eq:Cmat}
\mathbf{C}_\text{X} = \begin{pmatrix}
C_\text{X}^{+1}(t<0) & C_\text{X}^{+1}(t>0) \\
C_\text{X}^{-1}(t<0) & C_\text{X}^{-1}(t>0)
\end{pmatrix}, 
\end{equation}
which is defined by explicitly distinguishing between the positive and negative time-domains for the particle ($C_\text{X}^{+1}$) and hole ($C_\text{X}^{-1}$) correlation functions 
(explicit expressions are given in Table.~\ref{tab:ppf}). As illustrated in Fig.~\ref{fig:schematics}(b), the correlation matrix $\mathbf{C}_\text{X}$ for each type of purified pseudofermions has only one nonvanishing entry. We take full advantage of this unique feature to derive an easily implementable and accurate scheme for constructing effective models of fermionic reservoirs.
To demonstrate the accuracy, efficiency and potentials of our method, we benchmark our results by simulating the particle transport in a spinless fermionic chain, an example considered in Ref.~\cite{PhysRevX.10.031040}. In particular, our method provides improved numerical results because of its capability of capturing complex spectral structures alongside its optimized modeling of the reservoirs.

This paper is organized as follows. In Sec.~\ref{sec:model}, we describe the microscopic model of general fermionic open quantum systems and show that under appropriate assumptions the reduced system dynamics depends only on the free two-time reservoir correlations. In Sec.~\ref{sec:pf}, we review 
the pseudofermion approach~\cite{PhysRevResearch.5.033011}, which is the starting point of our method. In Sec.~\ref{sec:ppf}, we present the core results of this paper, namely, the definitions of four types of purified pseudofermions. In Sec.~\ref{sec:ppfmodel}, we then present explicitly a purified pseudofermion model based on exponential decompositions of the positive-time contributions of the reservoir correlation functions. In Sec.~\ref{sec:numerics}, we present numerical results to demonstrate the validity of our method and its capability in terms of numerical accuracy and efficiency. A summary of this paper is given in Sec.~\ref{sec:conclusion}.

\section{Fermionic Open Quantum Systems}\label{sec:model}

We start by defining the general model of fermionic open quantum systems consisting of a system $S$ and a fermionic reservoir $R$, described by the Hamiltonian
\begin{equation}
H = H_S + H_R + H_I, 
\end{equation}
where $H_S$ is the system Hamiltonian and  
\begin{equation}
H_R = \sum_k \epsilon_k c_k^\dagger c_k, 
\end{equation}
is the reservoir Hamiltonian, with $c_k$ ($c_k^\dagger$) destroying (creating) a reservoir fermion with energy $\epsilon_k$, 
satisfying the anticommutation relation $\{c_k,c_{k'}^\dagger\}=\delta_{kk'}$. The interaction Hamiltonian $H_I = \sum_k(g_ksc_k^\dagger + g_k^*c_k s^\dagger)$ describes the tunneling between the system and the reservoir where $s^\dagger$ ($s$) creates (destroys) a fermion in the system $S$ with $g_k$ the hopping parameters and $g_k^*$ their complex conjugate. 

We assume a factorized initial state $\rho(0)=\rho_S(0)\otimes \rho_R^\text{eq}$, where $\rho_S(0)$ is the system initial state and $\rho_R^\text{eq}=\exp(-\beta\sum_k(\epsilon_k-\mu)c_k^\dagger c_k)/Z$ denotes the reservoir thermal state at inverse temperature $\beta$ and chemical potential $\mu$. Here, $Z=\text{Tr}_R
\exp(-\beta\sum_k(\epsilon_k-\mu)c_k^\dagger c_k)$ is the partition function to ensure the normalization of $\rho_R^\text{eq}$. Our main interest is the reduced density matrix $\rho_S(t) = \text{Tr}_R\rho(t)$ obtained by tracing over all reservoir degrees of freedom in the full density matrix $\rho(t) = \exp(-iHt)\rho(0)\exp(iHt)$. In Ref.~\cite{PhysRevB.105.035121}, using a purely operator-based approach, some of us showed that $\rho_S(t)$ can be expressed in the following compact form~\cite{PhysRevB.105.035121,PhysRevResearch.5.033011}
\begin{equation}\label{eq:rhoS}
\rho_S(t) = \mathcal{T}_Se^{-\mathcal{F}[t,s,C^\sigma]}\rho_S(0), 
\end{equation}
where $\mathcal{T}_S$ ensures the time-ordering of system fermionic operators and the influence superoperator $\mathcal{F}[t,s,C^\sigma]$ encodes all environmental effects and can be explicitly written as~\cite{PhysRevB.105.035121,PhysRevResearch.5.033011}
\begin{equation}
\begin{array}{lll}
\mathcal{F}[\boldsymbol{\cdot}] &\displaystyle = \sum_{\sigma=\pm} \int_0^tdt_2\int_0^{t_2}dt_1 \Big( s^{\bar\sigma}(t_2)[\boldsymbol{\cdot}] - \mathcal{P}[\boldsymbol{\cdot}s^{\bar\sigma}(t_2)] \Big) \\
&\displaystyle \times \Big( C^\sigma(t_2-t_1) s^\sigma(t_1)[\boldsymbol{\cdot}] + C^{\bar\sigma,*}(t_2-t_1)\mathcal{P}[\boldsymbol{\cdot}s^\sigma(t_1)] \Big). 
\end{array}
\end{equation}
Here, $\sigma=\pm1$, $\bar\sigma=-\sigma$, and we have defined the operators $s^{\sigma=-1} \equiv s$ and $s^{\sigma=+1}\equiv s^\dagger$, and the parity superoperator
\begin{equation}
\mathcal{P}[\boldsymbol{\cdot}]  = \left(\prod_k e^{i\pi s_k^\dagger s_k}\right)[\boldsymbol{\cdot}]\left(\prod_k e^{-i\pi s_k^\dagger s_k}\right), 
\end{equation}
involving all fermions $s_k$ in the system $S$. In this model, the effects of the bath on the system are stationary. This is a consequence of the fact that the reservoir initial state is invariant under the free dynamics, i.e., $[H_R,\rho_R^\text{eq}]=0$. This further implies that the reservoir correlations $C^\sigma(t)\equiv C^\sigma(t_2-t_1)$ depend on the time difference  $t=t_2-t_1$ and they are explicitly expressed as
\begin{equation}\label{eq:Ct}
\begin{array}{lll}
\displaystyle C^\sigma(t) &\displaystyle = \text{Tr}_R\left[B^\sigma(t_2)B^{\bar\sigma}(t_1)\rho_R^\text{eq}\right] \\
&\displaystyle = \int_{-\infty}^\infty d\epsilon\frac{J(\epsilon)}{\pi} e^{i\sigma\epsilon t}\left[\frac{1-\sigma}{2}+\sigma f(\epsilon)\right], 
\end{array}
\end{equation}
where $B^{+1} = \sum_kg_kc_k$ and $B^{-1} = \sum_kg_kc_k^\dagger$ are the bath coupling operators, $f(\epsilon) = 1/(1+e^{\beta(\epsilon-\mu)})$ is the Fermi distribution function, and 
\begin{equation}\label{eq:J}
J(\epsilon) = \pi\sum_k g_k^2\delta(\epsilon-\epsilon_k), 
\end{equation}
denotes the reservoir hybridization function. From Eq.~(\ref{eq:Ct}), we can see that $C^\sigma(t)$ further satisfy the time-reversal relation
\begin{equation}\label{eq:trrelation}
C^\sigma(t) = C^{\sigma*}(-t), 
\end{equation}
which will be used later in our construction of the purified pseudofermion model. We note that the influence superoperator approach~\cite{PhysRevB.86.235432,PhysRevB.105.035121} employed here is equivalent to the well-known Feynman-Vernon influence functional formalism for fermionic reservoirs~\cite{FEYNMAN1963118,doi:10.1142/8334,PhysRevB.78.235311,Jin_2010,PhysRevB.105.125417}. In fact, the equivalence can be established explicitly by switching back to the Schr\"odinger picture in Eq.~(\ref{eq:rhoS}) and rephrasing the resulting propagator in the fermionic coherent-state representation as a path integral over Grassmann numbers~\cite{PhysRevB.78.235311,Jin_2010,PhysRevB.105.125417}.

Another physical quantity of practical interest is the particle (or charge) current flowing out of the reservoir, $I\equiv dN/dt$, defined as the time derivative of the reservoir particle number operator $N=\sum_kc_k^\dagger c_k$. Using the Heisenberg equation of motion $dN/dt = -i[N,H]$, the current can be further explicitly expressed as  
\begin{equation}\label{eq:currop}
I = -i (s B^{-1} - B^{+1}s^\dagger). 
\end{equation}
The reduced density matrix in Eq.~(\ref{eq:rhoS}) and the non-equilibrium particle current in Eq.~(\ref{eq:currop}) are the main quantities to be analyzed in this paper. 

\begin{widetext}

\begingroup
\setlength{\tabcolsep}{10pt} 
\renewcommand{\arraystretch}{1.5} 
\begin{table}[t]
\centering
\begin{tabular}{c || c | c } 
\toprule
PPF & Positive path: $\epsilon\to\epsilon+ia,~\gamma\to\gamma+a$  & Negative path: $\epsilon\to\epsilon-ia,~\gamma\to\gamma+a$  \\
\midrule
\multirow{4}{*}{$n=1$} & Type II: & Type I:  \\
& $\mathcal{L}_\text{II}[\boldsymbol{\cdot}] = (\epsilon-i\gamma)[\boldsymbol{\cdot}]dd^\dagger - \lambda[\boldsymbol{\cdot}](sd^\dagger + ds^\dagger) + \lambda s^\dagger[\boldsymbol{\cdot}]d$ & $\mathcal{L}_\text{I}[\boldsymbol{\cdot}] = -(\epsilon+i\gamma)dd^\dagger[\boldsymbol{\cdot}] + \lambda (sd^\dagger + ds^\dagger)[\boldsymbol{\cdot}] - \lambda d^\dagger[\boldsymbol{\cdot}]s$  \\
& $C_\text{II}^{+1}(t) = \lambda^2 e^{i\epsilon t + \gamma t}\Theta(-t)$,~$C_\text{II}^{-1}(t)=0$ & $C_\text{I}^{+1}(t)=\lambda^2 e^{i\epsilon t - \gamma t}\Theta(t)$,~$C_\text{I}^{-1}(t)=0$ \\
& $\rho_\text{II}(0)=|1\rangle\langle1|$ & $\rho_\text{I}(0)=|1\rangle\langle1|$ \\ 
\midrule
\multirow{4}{*}{$n=0$} & Type III: & Type IV: \\
& $\mathcal{L}_\text{III}[\boldsymbol{\cdot}] = (\epsilon-i\gamma)d^\dagger d[\boldsymbol{\cdot}] + \lambda (sd^\dagger + ds^\dagger)[\boldsymbol{\cdot}] + \lambda d[\boldsymbol{\cdot}]s^\dagger$ & $\mathcal{L}_\text{IV}[\boldsymbol{\cdot}] = -(\epsilon+i\gamma)[\boldsymbol{\cdot}]d^\dagger d - \lambda [\boldsymbol{\cdot}](sd^\dagger + ds^\dagger) - \lambda s[\boldsymbol{\cdot}]d^\dagger$  \\
& $C_\text{III}^{+1}(t)=0$,~$C_\text{III}^{-1}(t)=\lambda^2 e^{-i\epsilon t-\gamma t}\Theta(t)$ & $C_\text{IV}^{+1}(t)=0$,~$C_\text{IV}^{-1}(t)=\lambda^2 e^{-i\epsilon t + \gamma t}\Theta(-t)$ \\
& $\rho_\text{III}(0)=|0\rangle\langle0|$ & $\rho_\text{IV}(0)=|0\rangle\langle0|$ \\
\bottomrule
\end{tabular}
\caption{The generating superoperators $\mathcal{L}_\text{X}$, the purified pseudofermion correlations $C^{\pm1}_\text{X}(t)$, and the intial states $\rho_\text{X}(0)$ of the four types $\text{X}\in\{\text{I},\text{II},\text{III},\text{IV}\}$ of purified pseudofermions. $\Theta(\pm t)$ represent the Heaviside step functions. 
}\label{tab:ppf}
\end{table}    
\endgroup

\end{widetext}

\section{Pseudofermion Method}\label{sec:pf}

Before describing our method, in this section we review the core ideas underlying the pseudofermion approach~\cite{PhysRevResearch.5.033011}, which constitutes the starting point of our method. We first note that because of the vast degrees of freedom in macroscopic reservoirs, the spectral density $J(\epsilon)$ in Eq.~(\ref{eq:J}) is often assumed as a continuous function of the energy $\epsilon$. Such continuum is difficult to be accounted for in numerics and its treatment often distinguishes different methods. In the pseudofermion approach~\cite{PhysRevResearch.5.033011} and several other related ones~\cite{PhysRevLett.110.086403,PhysRevB.89.165105,PhysRevB.92.125145,PhysRevB.94.155142,PhysRevLett.121.137702,Dorda_2017,Chen_2019,PhysRevX.10.031040,PhysRevResearch.5.033011}, the original continuum is modeled as a fictitious environment consisting of discrete, damped ancillary fermions. Because of the Gaussianity of noninteracting reservoirs, it can be shown that this fictitious environment gives rise to identical environmental effects, provided that the equivalence condition, i.e., the two environments share the same reservoir correlation functions (see Eq.~(\ref{eq:Cequiv}) below), is satisfied. Importantly, the pseudofermion approach 
admits complex model parameters which, on one hand, permits more freedom in constructing the effective model and, on the other hand, constitutes the defining feature of the pseudofermion approach, distinguishing it from other related methods~\cite{PhysRevLett.110.086403,PhysRevB.89.165105,PhysRevB.92.125145,PhysRevB.94.155142,PhysRevLett.121.137702,Dorda_2017,Chen_2019,PhysRevX.10.031040,PhysRevResearch.5.033011}.

We now proceed to review further details of the pseudofermion approach. Specifically, a pseudofermion is defined as a fermion $d$ which can tunnel into a background reservoir characterized by a flat band structure such that its effects can be modeled by a Markovian dissipator with an effective Markovian damping with rate $\gamma$ and an average thermal occupation $n$. As a consequence, when the pseudofermion $d$ is further coupled to the system $S$ via the hopping interaction $\lambda(sd^\dagger + ds^\dagger)$ in terms of a hopping parameter $\lambda$, the state $\rho_\text{pf}(t)$ of the composite system $S+d$ obeys the following Lindblad master equation 
\begin{equation}\label{eq:pfeom}
\begin{array}{lll}
\displaystyle i\frac{d\rho_\text{pf}(t)}{dt} &\displaystyle = [H_S + \epsilon d^\dagger d+ \lambda(s d^\dagger + d s^\dagger), \rho_\text{pf}] \\
&\displaystyle~~~ + i\gamma(1-n)(2d\rho_\text{pf} d^\dagger - d^\dagger d\rho_\text{pf} - \rho_\text{pf} d^\dagger d) \\
&\displaystyle~~~ + i\gamma n(2d^\dagger\rho_\text{pf} d - dd^\dagger\rho_\text{pf} - \rho_\text{pf} dd^\dagger), 
\end{array}
\end{equation}
where $\epsilon$ is the energy of the pseudofermion. The initial state of this composite system should take $\rho_\text{pf}(0)=\rho_S(0)\otimes\rho_d$ with $\rho_d=(1-n)|0\rangle\langle0|+n|1\rangle\langle1|$. 

The free correlation of this pseudofermion, defined when the system $S$ is decoupled, takes the analytical form
\begin{equation}\label{eq:Cpf}
C_\text{pf}^\sigma(t) = \lambda^2\left[\frac{1-\sigma}{2} + \sigma n\right] e^{i\sigma\epsilon t - \gamma\lvert t\rvert},   
\end{equation}
which corresponds to the contribution that this artificial environment consisting of one single damped fermion can offer in order to model the full correlation for the physical environment in Eq.~(\ref{eq:Ct}).
As pointed out in Ref.~\cite{PhysRevResearch.5.033011}, an important property of pseudofermions is that all of their parameters, including $\epsilon$, $\gamma$, $\lambda$ and $n$, can be analytically continued to the whole complex plane $\mathbb{C}$ to produce a larger class of pseudofermion correlation ans\"atz~(\ref{eq:Cpf}). In fact, this is the defining feature of the pseudofermion method~\cite{PhysRevResearch.5.033011} which distinguishes it from the auxiliary master equation approach in Refs.~\cite{PhysRevLett.110.086403,PhysRevB.89.165105,PhysRevB.92.125145,PhysRevB.94.155142,PhysRevLett.121.137702,Dorda_2017,Chen_2019,PhysRevX.10.031040,PhysRevResearch.5.033011}. 

In order to model more complex reservoir correlations, we simply introduce $N_\text{pf}$ pseudofermions, independently coupled to the system $S$, to satisfy (at least approximately) the equivalence condition 
\begin{equation}\label{eq:Cequiv}
C^\sigma(t) \cong \sum_{j=1}^{N_\text{pf}} C_{\text{pf},j}^\sigma(t), 
\end{equation}
with $C_{\text{pf},j}^\sigma(t)$ representing the free correlations of the $j$th pseudofermion $d_j$. This equivalence condition~(\ref{eq:Cequiv}) ensures that the reduced density matrix of the open-system model in Sec.~\ref{sec:model} can be reproduced by tracing out all pseudofermion degrees of freedom in $\rho_\text{pf}(t)$, i.e., $\rho_S(t) = \text{Tr}_\text{pf}\rho_\text{pf}(t)$. 

In addition to the reduced state of the system, non-equilibrium particle current can also be evaluated using the pseudofermion counterpart of the reservoir current operator~(\ref{eq:currop}), defined as 
\begin{equation}\label{eq:pfcurrentop}
I_\text{pf} = -i \sum_{j=1}^{N_\text{pf}} \lambda_j(sd_j^\dagger  - d_js^\dagger). 
\end{equation}
It is important to note that because $\lambda_j$ could be complex-valued, namely $\lambda_j\in\mathbb{C}$, $I_\text{pf}$ could be non-Hermitian. However, its expectation should be real and it satisfies $\text{Tr}[I\rho(t)] = \text{Tr}_{S+\text{pf}}[I_\text{pf}\rho_\text{pf}(t)]$, as guaranteed by the equivalence condition~(\ref{eq:Cequiv}). 

The pseudofermion approach~\cite{PhysRevResearch.5.033011} provides one of the most general theoretical frameworks for modeling fermionic reservoirs via ancillary degrees of freedom. Its generality is reflected in the fact that, all pseudofermion parameters, including the energy $\epsilon$, decay rate $\gamma$, coupling amplitude $\lambda$, and  average occupation number $n$, can in principle take complex values, i.e., $\lambda,\epsilon,\gamma,n\in\mathbb{C}$. Despite this flexibility, constructing an effective reservoir model with pseudofermions may remain a challenging task in many practical situations, especially when the spectral density $J(\epsilon)$ and/or the Fermi function $f(\epsilon)$ exhibit nontrivial structures, as commonly encountered near band edges or at low temperatures. In such cases, a large number of pseudofermions, i.e., $N_\text{pf}\gg1$, is often required to faithfully satisfy the equivalence condition~(\ref{eq:Cequiv}). As a result, the $4N_\text{pf}$ free parameters in the corresponding pseudofermion model must be searched in the space $\mathbb{R}^{8N_\text{pf}}$, rendering conventional optimization algorithms inefficient and even impractical. 

In the remainder of this work, we address this challenge by (i) introducing a purified version of pseudofermions in Sec.~\ref{sec:ppf}, and (ii) describing a procedure to construct effective models using purified pseudofermions in Sec.~\ref{sec:ppfmodel}. The key advantage of this purified formulation is that it allows the positive-time and
negative-time contributions of the reservoir correlation functions, denoted respectively as $C^\sigma(t>0)$ and $C^\sigma(t<0)$, to be modeled independently. This decoupling makes the purified version compatible with a number of existing decomposition algorithms~\cite{doi:10.1137/16M1106122,10.1063/5.0209348}. These key features ultimately enable the efficient and accurate construction of effective reservoir models.

\begin{figure}
\includegraphics[clip,width=8.5cm]{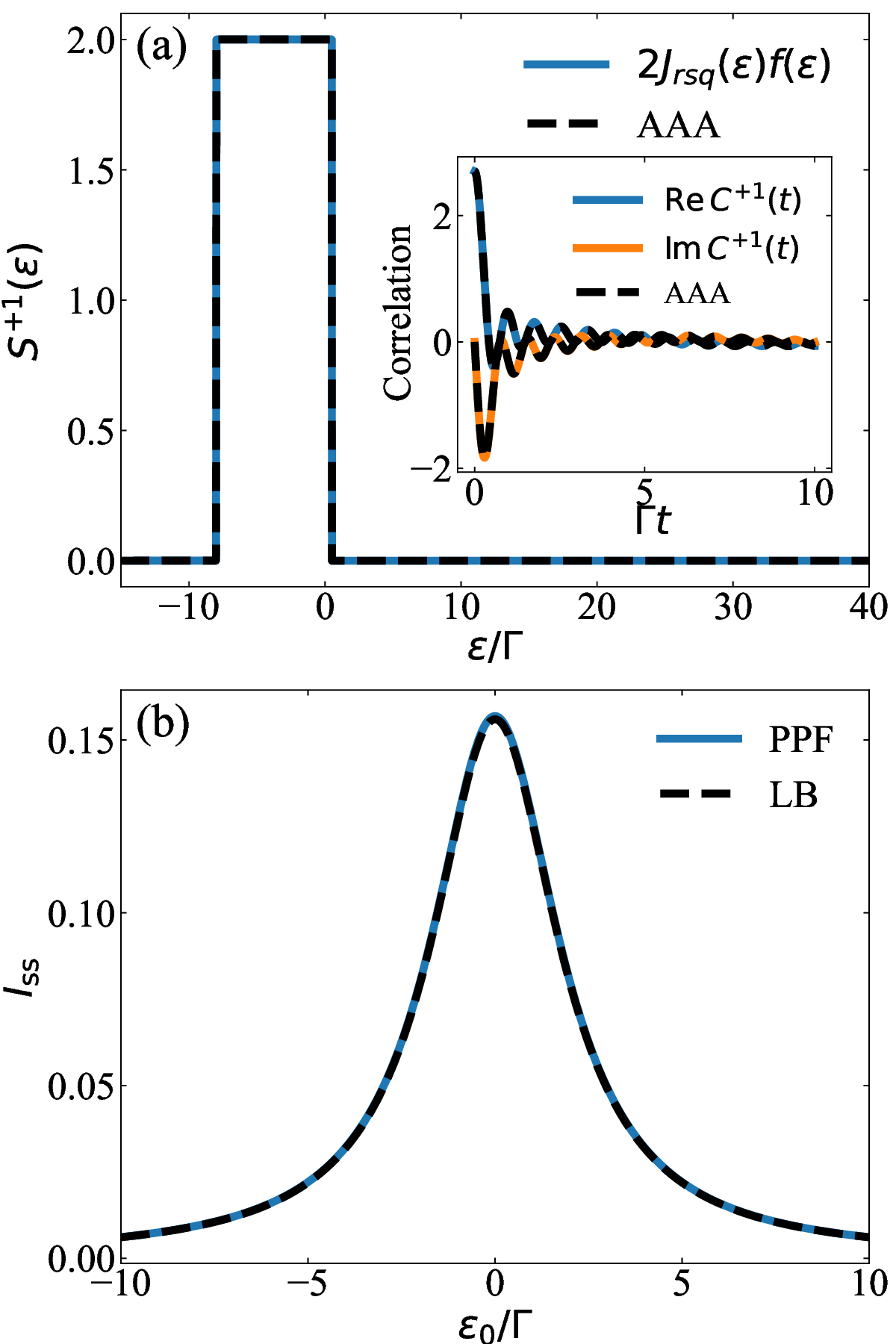}
\caption{In (a) the spectrum $S^{+1}(\epsilon)=2J_\text{rsq}(\epsilon)f(\epsilon)$ and the reservoir correlation $C^{+1}(t)$ (inset) of the left lead are plotted. Black dashed lines correspond to their approximations obtained with the AAA algorithm. In (b) the steady-state particle current $I_{ss}$ from the purified pseudofermion model (labeled as "PPF") is plotted as a function of the impurity energy $\epsilon_0$. Results of the LB formula is shown in black dashed line. Simulation parameters are: $W=8\Gamma$, $A=200$, $\mu_L=-\mu_R=W/16$, and $T_L=T_R=0$.
}\label{fig:impurity}
\end{figure}

\section{Purified Pseudofermions}\label{sec:ppf}

In this section, we present the core results of this paper, namely, the definitions of four types of purified pseudofermions, as listed in Table~\ref{tab:ppf}. As mentioned,  these new objects are obtained by performing the analytical continuation of the frequency $\epsilon$ and decay rate $\gamma$ of a pseudofermion, with its occupation number $n$ being either $n=0$ or $n=1$, along one of the following paths in the complex plane
\begin{equation}\label{eq:acpath}
\epsilon \to \epsilon \pm ia,~~~\gamma \to \gamma + a, 
\end{equation}
with $a\in\mathbb{R}_+$ a free frequency parameter, in the limit $a\to+\infty$. Depending on the possible values ($0$ or $1$) of the occupation number $n$ and the possible analytical continuation paths given in Eq.~(\ref{eq:acpath}), four distinct cases can be identified: (i) $n=1$, $\epsilon \to \epsilon+ia$, $\gamma \to \gamma+a$; (ii)  $n=1$, $\epsilon \to \epsilon-ia$, $\gamma \to \gamma+a$; (iii)  $n=0$, $\epsilon \to \epsilon+ia$, $\gamma \to \gamma+a$; (iv) and $n=0$, $\epsilon \to \epsilon-ia$, $\gamma \to \gamma+a$. In the remaining part of this section we focus on the analysis of the case (i), corresponding to the type II purified pseudofermion, and present the detailed derivations for the other cases (ii-iv) in Appendix~\ref{app:ppf}. 

As in the analysis of purified pseudomodes~\cite{liang2024purifiedinputoutputpseudomodemodel}, we begin by mapping the pseudofermion equation of motion in Eq.~(\ref{eq:pfeom}) to a Schr\"odinger-type equation in the superfermion representation~\cite{10.1063/1.3548065}, which is achieved by defining the following {\it left-vacuum} ket vector~\cite{PhysRevX.10.031040,10.1063/1.3548065} 
\begin{equation}
|I\rangle = \sum_{m,n=0,1} (s^\dagger)^m(\tilde{s}^\dagger)^m(d^\dagger)^n(\tilde{d}^\dagger)^n|0\rangle, 
\end{equation}
where we have introduced two ancillary $\tilde{s}$ and $\tilde{d}$ fermions, and $|0\rangle$ denotes the vacuum. Note that we have neglected other fermions in the system $S$ as their presence has no influence on the results presented later in this section. Useful properties of $|I\rangle$ are the following conversion rules
\begin{equation}
\begin{array}{lll}
&\displaystyle s^\dagger |I\rangle = - \tilde{s}|I\rangle,~~~s|I\rangle = \tilde{s}^\dagger |I\rangle, \\
&\displaystyle d^\dagger |I\rangle = - \tilde{d}|I\rangle,~~~d|I\rangle = \tilde{d}^\dagger |I\rangle. 
\end{array}
\end{equation}
We can now apply the master equation~(\ref{eq:pfeom}) to $|I\rangle$ to write
\begin{equation}\label{eq:pfeoml}
i\frac{d|\rho\rangle}{dt} =  L|\rho\rangle = (L_S+L_\text{pf})|\rho\rangle, 
\end{equation}
in terms of the vectorized density matrix $|\rho\rangle = \rho|I\rangle$, where $L_S = H_S - \tilde{H}_S$. Here, $\tilde{H}_S$ is defined by satisfying the equality $\tilde{H}_S|I\rangle = H_S|I\rangle$ such that $\tilde{H}_S$ contains the operators $\tilde{s}$ and $\tilde{s}^\dagger$ only. 
The generator $L_\text{pf}$ is defined as
\begin{equation}\label{eq:pfL}
\begin{array}{lll}
L_\text{pf} &\displaystyle = \epsilon\big(d^\dagger d - \tilde{d}^\dagger\tilde{d} \big) + \lambda\big(sd^\dagger+ds^\dagger - \tilde{s}\tilde{d}^\dagger - \tilde{d}\tilde{s}^\dagger)  \\
&\displaystyle~~~ - i\gamma(1-n)\big( 2d\tilde{d} + d^\dagger d + \tilde{d}^\dagger \tilde{d} \big) \\
&\displaystyle~~~ + i\gamma n\big( 2d^\dagger\tilde{d}^\dagger - dd^\dagger - \tilde{d}\tilde{d}^\dagger \big).    
\end{array} 
\end{equation}
Accordingly, the initial state becomes $|\rho(0)\rangle = |\rho_S(0)\rangle\otimes|\rho_d\rangle$ with $|\rho_d\rangle = (1-n)|0\tilde{0}\rangle + n|1\tilde{1}\rangle$. We emphasize that vectorizing the master equation~(\ref{eq:pfeom}) in the superfermion representation~\cite{10.1063/1.3548065} is equivalent to the Choi isomorphism~\cite{Nielsen_Chuang_2010} of quantum channels.

Along the analytical continuation path specified in the case (i), the generator in Eq.~(\ref{eq:pfL}) simplifies as
\begin{equation}\label{eq:ac1}
\begin{array}{lll}
L_{\text{pf},a}
&\displaystyle = -(\epsilon + i(\gamma+2a))dd^\dagger + (\epsilon - i\gamma)\tilde{d}\tilde{d}^\dagger \\
&\displaystyle~~~ + 2i(\gamma+a)d^\dagger\tilde{d}^\dagger  + \lambda\big(sd^\dagger+ds^\dagger - \tilde{s}\tilde{d}^\dagger - \tilde{d}\tilde{s}^\dagger \big),  
\end{array} 
\end{equation}
where we have added $a$ in the subscript of $L_\text{pf}$ to emphasize its dependence on the free parameter $a$ along the continuation path, and where the fermions $d$ and $\tilde{d}$ now both evolve from their occupied state $|\rho_d\rangle = |1\tilde{1}\rangle$. A key observation  is that, in Eq.~(\ref{eq:ac1}), the decay rate of the $d$ fermion is given by $\gamma+2a$, implying that any transition out of its initial occupied state is strongly suppressed and even completely prohibited in the limit $a\to+\infty$. In other words, in this limit, the $d$ fermion always stays in its initial occupied state during the evolution. As a consequence, it becomes irrelevant to the dynamics and it can be dynamically decoupled from the rest degrees of freedom. This is the intuitive picture underlying the idea of purified pseudofermions. 

To confirm this intuitive argument, we now show that the interaction terms in Eq.~(\ref{eq:ac1}) that cause transitions out of the occupied state of the $d$ fermion can be adiabatically eliminated. Following the general ideas used in 
Ref.~\cite{liang2024purifiedinputoutputpseudomodemodel}, we proceed to decompose the generator $L_{\text{pf},a}$ as $L_{\text{pf},a} = L_\text{f} + L_0 + V$, where the three components are defined as 
\begin{equation}\label{eq:ac1LfL0V}
\begin{array}{lll}
L_\text{f} &\displaystyle = (\epsilon-i\gamma)\tilde{d}\tilde{d}^\dagger - \lambda (\tilde{s}\tilde{d}^\dagger + \tilde{d}\tilde{s}^\dagger), \\
L_0 &\displaystyle = -(\epsilon+i(\gamma+2a))dd^\dagger + 2i(\gamma+a)d^\dagger\tilde{d}^\dagger + \lambda sd^\dagger, \\
V &\displaystyle = \lambda ds^\dagger. 
\end{array}
\end{equation}
Our goal is to look for an operator $S$ of order $\mathcal{O}(a^{-1})$ such that $V$ can be eliminated in $L_{\text{pf},a}$ by the Schrieffer-Wolff (SW) transformation 
\begin{equation}\label{eq:SW}
e^{S}L_{\text{pf},a}e^{-S} = L_{\text{pf},a} + [S,L_{\text{pf},a}] + \frac{1}{2!}[S,[S,L_{\text{pf},a}]] + \cdots. 
\end{equation}
A convenient solution is to satisfy the following operator equation
\begin{equation}\label{eq:Seq}
[S,L_\text{f}+L_0] = -V + V'+\mathcal{O}(a^{-1}),  
\end{equation}
where the operator $V'$ represents the emerging interaction mediated by the eliminated mode. Importantly, we require $V'$ not to contain $d$, and to be of order $\mathcal{O}(a^0)$. This condition ensures nested commutators appearing in the SW transformation in Eq.~(\ref{eq:SW}), such as $[S,\cdots,[S,L_\text{f}+L_0]]$ with at least two $S$, or $[S,\cdots,[S,V]]$ with at least one $S$, to be at least of order $\mathcal{O}(a^{-1})$. Hence, in the limit $a\to+\infty$, they vanish in the expansion of the SW transformation in Eq.~(\ref{eq:SW}), leading to the simple equation 
\begin{equation}
\lim_{a\to+\infty}e^{S}L_{\text{pf},a}e^{-S} = L_S + L_\text{f}+L_0+V',   
\end{equation}
where the term $V$ in Eq.~(\ref{eq:ac1LfL0V}) is eliminated, as promised. Since by construction none of the operators $L_S$, $L_\text{f}$, $L_0$ [see their definitions in Eq.~(\ref{eq:ac1LfL0V})] and $V'$ [as we require $V'$ not to end the operator $d$] trigger the transition out of the occupied state of the $d$ fermion, we conclude that the $d$ fermion remains in its initial occupied state so that $L_0$ can be discarded and the evolution of the rest degrees of freedom is generated by $L_S+L_{\text{pf},\infty}$ with 
\begin{equation}\label{eq:newGen}
L_{\text{pf},\infty} \equiv L_\text{f}+V'. 
\end{equation}

For the case (i) analyzed here, the following operator
\begin{equation}
S = \frac{1}{2ia} \lambda s^\dagger d, 
\end{equation}
satisfies the equation 
\begin{equation}
[S, L_\text{f}+L_0] = -\lambda ds^\dagger + \lambda s^\dagger\tilde{d}^\dagger + \mathcal{O}(a^{-1}),  
\end{equation}
where the first term on the right-hand side cancels out $V$ in Eq.~(\ref{eq:ac1LfL0V}) exactly and the new mediated interaction term is $V' = \lambda s^\dagger\tilde{d}^\dagger$. The corresponding effective dynamical generator is 
\begin{equation}
L_\text{II} \equiv L_{\text{pf},\infty} = (\epsilon-i\gamma)\tilde{d}\tilde{d}^\dagger - \lambda(\tilde{s}\tilde{d}^\dagger + \tilde{d}\tilde{s}^\dagger) + \lambda s^\dagger\tilde{d}^\dagger. 
\end{equation}
In other words, in the original density matrix representation, the type II purified pseudofermion corresponding to the case (i) obeys the equation of motion 
\begin{equation}
\label{eq:typeI}
    id\rho/dt = [H_S,\rho] + \mathcal{L}_\text{II}\rho\;,
\end{equation}
where the generating superoperator is
\begin{equation}
\label{eq:LindtypeI}
\mathcal{L}_\text{II}[\boldsymbol{\cdot}] = (\epsilon-i\gamma)[\boldsymbol{\cdot}]dd^\dagger - \lambda[\boldsymbol{\cdot}](sd^\dagger+ds^\dagger) + \lambda s^\dagger[\boldsymbol{\cdot}]d,   
\end{equation}
and the initial state is 
\begin{equation}
    \label{eq:initial}
    \rho_S(0)\otimes\rho_\text{II}(0)\;,
\end{equation}
with $\rho_\text{II}(0)=|1\rangle\langle1|$. 

Consistent with the above analysis, we can further check that, by setting $n=1$ and taking the limit $a\to+\infty$ in the pseudofermion correlation ans\"atz~(\ref{eq:Cpf}), we obtain the correlations of the type II purified pseudofermion as
\begin{equation}\label{eq:corrp1}
\begin{array}{lll}
C_\text{II}^{+1}(t) &\displaystyle = \lim_{a\to+\infty} \lambda^2 e^{-i(\epsilon+ia)t - (\gamma+a)\lvert t\rvert} \\
&\displaystyle = \lambda^2 e^{i\epsilon t - \gamma \lvert t\rvert}\Theta(-t), 
\end{array} 
\end{equation}
and 
\begin{equation}\label{eq:corrm1}
C_\text{II}^{-1}(t) = 0.    
\end{equation}
This concludes the analysis for the type II purified pseudofermions. We highlight that, despite being inspired by adiabatic elimination techniques, the equation of motion in Eq.~(\ref{eq:typeI}) encodes, without approximations, the full influence of a pseudofermion whose correlations take the unusual form in Eqs.~(\ref{eq:corrp1}) and (\ref{eq:corrm1}).

Before closing this rather technical section, we summarize the main results obtained so far. By using a vectorized formalism, we showed that the left action of the pseudofermion $d$ in Eq.~(\ref{eq:pfeoml}) can be adiabatically eliminated so that its effects on the system can be identically encoded in the  Lindblad operator in Eq.~(\ref{eq:LindtypeI}) with initial state given by Eq.~(\ref{eq:initial}). This can be also interpreted as the dynamical consequence of the fact that the corresponding free correlation is non-trivial only at negative times, see Eq.~(\ref{eq:corrp1}). As shown in Appendix~\ref{app:ppf} and summarized in Table~\ref{tab:ppf}, the results in Eqs.~(\ref{eq:LindtypeI}-\ref{eq:corrm1}) can be readily generalized to also describe the remaining classes $\text{X}=$ I, III, and IV, i.e. to describe the corresponding purified pseudofermions modeling the effects encoded in the correlation $C_\text{X}^{\pm1}(t)$ in terms of their generating superoperator $\mathcal{L}_\text{X}$, and  initial state $\rho_\text{X}(0)$.

\section{Purified Pseudofermion Model for Fermionic Reservoirs}\label{sec:ppfmodel}

As sketched in Fig.~\ref{fig:schematics}(a), we are now ready to explicitly construct an effective model, consisting of $N_\text{X}$ purified pseudofermions for each type $\text{X}\in\{\text{I},\text{II},\text{III},\text{IV}\}$, such that the following equivalence condition is satisfied,
\begin{equation}\label{eq:equivcondition}
C^\sigma(t) \cong C_\text{ppf}^\sigma (t) \equiv \sum_{\text{X}}\sum_{l=1}^{N_\text{X}}C_{\text{X},l}^\sigma(t), 
\end{equation}
where $C_{\text{X},l}^\sigma(t)$ denotes the correlations of the $l$th type X purified pseudofermion, as listed in Table~\ref{tab:ppf}. Similarly with the original pseudofermion theory, this equivalence condition ensures that this effective model is capable of reproducing the reduced density matrix $\rho_S(t)$ in Eq.~(\ref{eq:rhoS}) and the nonequilibrium particle current $I$ in Eq.~(\ref{eq:currop}) of the original open model defined in Sec.~\ref{sec:model}. 

The remaining task is to define the free parameters appearing in $C_{\text{X},l}^\sigma(t)$, including the coupling amplitudes $\lambda_{\text{X},l}$, energies $\epsilon_{\text{X},l}$ and decay rates $\gamma_{\text{X},l}$. To this end, we can  employ the AAA algorithm~\cite{doi:10.1137/16M1106122,10.1063/5.0209348} to generate the following decompositions in terms of exponentials
\begin{equation}\label{eq:decomppos}
C^\sigma(t>0) \approx \sum_{l=1}^{N_\sigma} w_l^\sigma e^{i\sigma\epsilon_l^\sigma t - \gamma_l^\sigma t}, 
\end{equation}
where the parameters $w_l^\sigma\in\mathbb{C}$, $\epsilon_l^\sigma\in\mathbb{R}$, and $\gamma_l^\sigma\in\mathbb{R}_+$ are determined during the decomposition procedure, and $N_\sigma$ is the number of exponentials. Recall that $C^\sigma(t>0)$ represent the positive-time contributions of the correlation functions $C^\sigma(t)$ in Eq.~(\ref{eq:Ct}). To keep the presentation concise, we postpone the description of the AAA algorithm to the end of this section. Here, we highlight that this kind of exponential decompositions can in principle be made arbitrarily accurate by increasing $N_\sigma$, and can also be realized via several other algorithms, see Ref.~\cite{10.1063/5.0209348} and references therein. In fact, these algorithms have been successfully employed within the HEOM approach to extend its capability beyond the high-temperature regime. As we demonstrate here, their application within our purified pseudofermion theory is also going to bring significant advantages in the modeling and simulation of fermionic reservoirs. 

Using the time-reversal property in Eq.~(\ref{eq:trrelation}), the negative-time contributions $C^\sigma(t<0)$ are decomposed accordingly as
\begin{equation}\label{eq:decompneg}
C^\sigma(t<0) \approx \sum_{l=1}^{N_\sigma} w_l^{\sigma*} e^{i\sigma\epsilon_l^\sigma t + \gamma_l^\sigma t}. 
\end{equation}
Combining the two decompositions in Eqs.~(\ref{eq:decomppos}) and (\ref{eq:decompneg}), we conclude that $C^\sigma(t)$ are decomposed as 
\begin{equation}\label{eq:decompfull}
C^\sigma(t) \approx \sum_{l=1}^{N_\sigma} w_l^\sigma e^{i\sigma\epsilon_l^\sigma t - \gamma_l^\sigma t}\Theta(t) + \sum_{l=1}^{N_\sigma} w_l^{\sigma*} e^{i\sigma\epsilon_l^\sigma t + \gamma_l^\sigma t}\Theta(-t), 
\end{equation}
where the Heaviside step function $\Theta(\pm t)$ ensure consistency with the two decompositions in Eqs.~(\ref{eq:decomppos}) and (\ref{eq:decompneg}).

Finally, by comparing Eq.~(\ref{eq:decompfull}) with the ansatz $C_\text{ppf}^\sigma(t)$ in Eq.~(\ref{eq:equivcondition}), we can identify a total of $N_\text{ppf} = 2(N_+ + N_-)$ purified pseudofermions distributed as
$N_\text{I} = N_\text{II} = N_+$ and $N_\text{III} = N_\text{IV}  = N_-$. Their parameters are given by
\begin{equation}
\begin{array}{lll}
&\displaystyle \text{Type I: }  \lambda_{\text{I},l} = \sqrt{w_l^{+}},\epsilon_{\text{I},l} = \epsilon_l^+,\gamma_{\text{I},l} = \gamma_l^+; \\
&\displaystyle \text{Type II: } \lambda_{\text{II},l} = \sqrt{w_l^{+*}},\epsilon_{\text{II},l} = \epsilon_l^+,\gamma_{\text{II},l} = \gamma_l^+; \\
&\displaystyle \text{Type III: }  \lambda_{\text{III},l} = \sqrt{w_l^{-}},\epsilon_{\text{III},l} = \epsilon_l^-,\gamma_{\text{III},l} = \gamma_l^-; \\
&\displaystyle \text{Type IV: }  \lambda_{\text{IV},l} = \sqrt{w_l^{-*}},\epsilon_{\text{IV},l} = \epsilon_l^-,\gamma_{\text{IV},l} = \gamma_l^-. 
\end{array}
\end{equation}
This completes our construction of the purified pseudofermion model. 

These parameters provide an explicit form to their 
corresponding generators $\mathcal{L}_\text{X}$ in Table~\ref{tab:ppf}. This allows us to  conclude that the state $\rho_\text{ppf}(t)$ describing the system $S$ and all purified pseudofermions obeys the following Lindblad-like master equation 
\begin{equation}\label{eq:ppfeom}
\begin{array}{lll}
\displaystyle i\frac{d\rho_\text{ppf}}{dt} &\displaystyle = [H_S,\rho_\text{ppf}] + \sum_\text{X}\sum_{l=1}^{N_\text{X}} \mathcal{L}_{\text{X},l}[\rho_\text{ppf}], 
\end{array}
\end{equation}
with initial condition give by
\begin{equation}\label{eq:ppfini}
\rho_\text{ppf}(0) = \rho_S(0)\otimes \big(\prod_{l=1}^{N_+}d_{\text{I},l}^\dagger d_{\text{II},l}^\dagger\big) |0\rangle\langle0| \big(\prod_{l=1}^{N_+}d_{\text{III},l} d_{\text{IV},l}\big). 
\end{equation}
Ultimately, the reduced density matrix $\rho_S(t)$ is obtained by tracing out all purified pseudofermion degrees of freedom, i.e., $\rho_S(t) = \text{Tr}_\text{ppf}[\rho_\text{ppf}(t)]$. In addition, according to Eq.~(\ref{eq:pfcurrentop}), the particle current operator in the purified pseudofermion model is explicitly written as 
\begin{equation}\label{eq:ppfcurrentop}
\begin{array}{lll}
\displaystyle I_\text{ppf} &\displaystyle = -i \sum_{\text{X}}\sum_{l=1}^{N_\text{X}}  \lambda_\text{X}(sd_{\text{X},l}^\dagger  - d_{\text{X},l}s^\dagger) .  
\end{array}
\end{equation}
As a consequence, Eqs.(\ref{eq:ppfeom}-\ref{eq:ppfcurrentop}) allow to track both the evolution of the reduced density matrix $\rho_S(t)$ and the nonequilibrium particle current $I(t)$. 

Before closing this section, we provide implementation details for the AAA algorithm~\cite{doi:10.1137/16M1106122,10.1063/5.0209348}, which we have employed in numerical simulations to generate spectral decompositions in the form of Eq.~(\ref{eq:decomppos}). We refer to Ref.~\cite{10.1063/5.0209348} for details about the implementation of other related algorithms. 

To apply the AAA algorithm, we consider the power spectra
\begin{equation}
S^\sigma(\epsilon) = \int_{-\infty}^{+\infty} dt C^\sigma(t) e^{-i\sigma\epsilon t} = 2J(\epsilon)\left[\frac{1-\sigma}{2}+\sigma f(\epsilon)\right], 
\end{equation}
defined as the Fourier transform of the reservoir correlation functions $C^\sigma(t)$. Generally speaking, this algorithm aims to find the rational approximation of a function defined on the real axis, such as $S^\sigma(\epsilon)$, in the barycentric representation
\begin{equation}\label{eq:SAAA}
S^\sigma_\text{AAA}(\epsilon) = \sum_{l=1}^M \frac{z^\sigma_lS^\sigma(\epsilon_l)}{\epsilon - \epsilon_l} \Big/ \sum_{l=1}^M \frac{z^\sigma_l}{\epsilon - \epsilon_l}, 
\end{equation}
such that $S^\sigma(\epsilon) \cong S^\sigma_\text{AAA}(\epsilon)$ for all $\epsilon\in\mathbb{R}$. Here, $M\ge1$ is an integer representing the order of the rational appproximation, $z^\sigma_l\in\mathbb{C}$ are free parameters, and $\{\epsilon_l \in\mathbb{R}| l = 1,\cdots,M\}$ is the set of support points where
$S^\sigma(\epsilon_l) = S^\sigma_\text{AAA}(\epsilon_l)$ by residue theorem. Once the barycentric representation $S^\sigma_\text{AAA}(\omega)$ is obtained, its poles $p_l^\sigma$ in the lower (for $\sigma=+1$) or upper (for $\sigma=-1$) half complex plane and the corresponding residues $r_l^\sigma$ can be solved to yield the spectral decomposition in Eq.~(\ref{eq:decomppos}), with the parameters given by 
\begin{equation}
w_l^\sigma = i\sigma r_l^\sigma,~\nu_l^\sigma = \text{Re}\,p_l^\sigma,~\gamma_l^\sigma = \sigma\text{Im}\,p_l^\sigma.
\end{equation}
In practice, one could first choose a relatively large support set, $M\gg1$, and post-selects the most important $N_\sigma$ poles by their contributions to the approximation in Eq.~(\ref{eq:SAAA}), until a preset error tolerance is reached.

\begin{figure}
\includegraphics[clip,width=8.5cm]{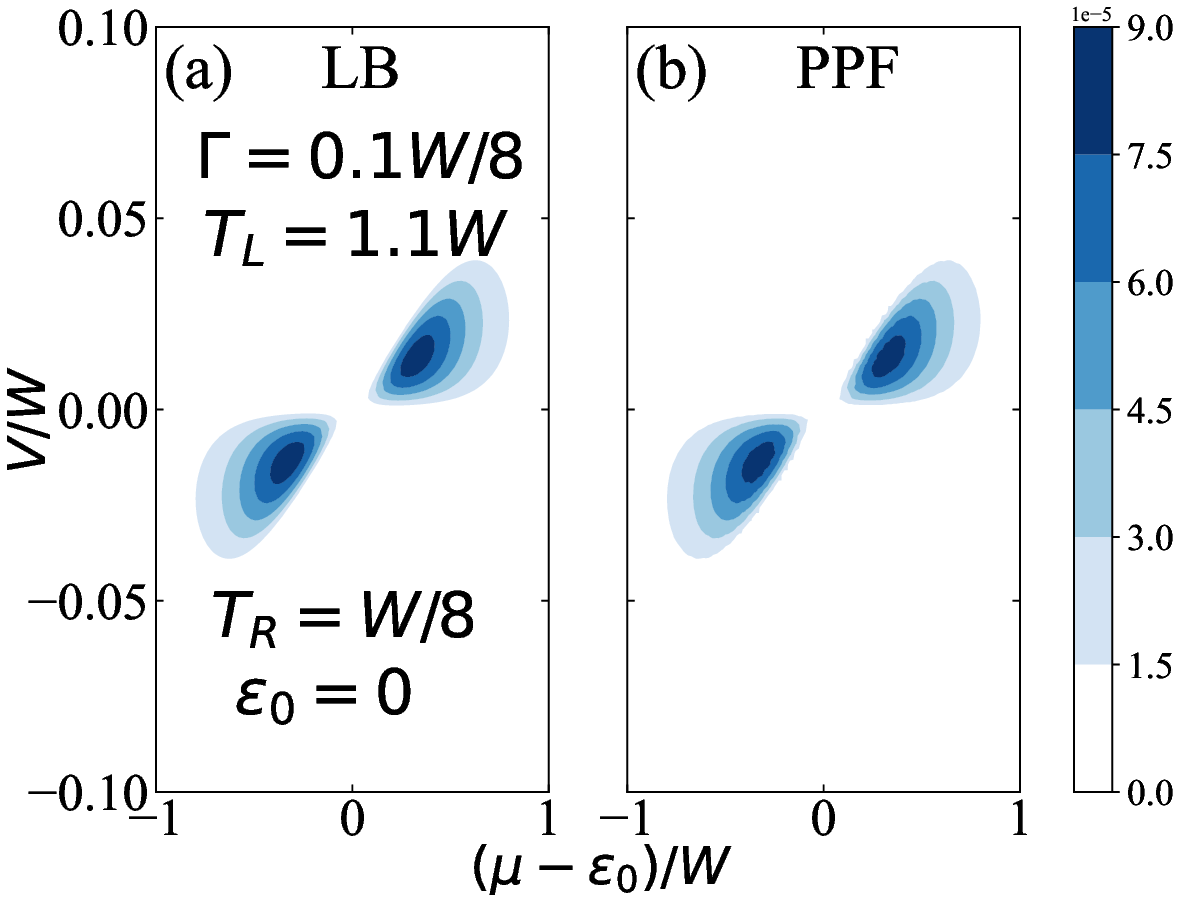}
\caption{The steady-state particle current $I_{ss}$ is plotted as a function of the average chemical potential $\mu=(\mu_L+\mu_R)/2$ and the bias $V=\mu_L-\mu_R$, using (a) the LB formula and (b) the purified pseudofermion model. In the region with $VI_{ss}<0$, we set $I_{ss}=0$ by hand. Simulation parameters are the same as those in Fig.~8 in Ref.~\cite{PhysRevX.10.031040}. 
}\label{fig:power}
\end{figure}

\section{Numerical demonstrations}\label{sec:numerics}

In order to demonstrate the validity, accuracy and efficiency of our method, in this section we consider the particle transport along a chain with $D$ sites able to host spinless fermions. The full model, including the chain and the two leads coupled, respectively, to the first ($j=1$) and the last ($j=D$) sites of the chain, is described by the Hamiltonian
\begin{equation}\label{eq:Hchain}
H = H_S + \sum_{\alpha,k}\epsilon_{\alpha k}c_{\alpha k}^\dagger c_{\alpha,k} + \sum_{\alpha,k}g_{\alpha k}(f_\alpha c_{\alpha k}^\dagger + c_{\alpha k}f_\alpha^\dagger), 
\end{equation}
where $c_{\alpha k}$ and $c_{\alpha k}^\dagger$ are the destroying and creation operators in the $\alpha=L$ and $R$ lead, while $\epsilon_{\alpha k}$ and $g_{\alpha k}$ denote the corresponding fermionic energy and hopping amplitudes, respectively. The left (right) lead $L$ ($R$) couples to the first (last) site of the chain, so that $f_L=f_1$ ($f_R=f_D$). The system Hamiltonian is
\begin{equation}\label{eq:chainHS}
H_S = \sum_{j=1}^D \epsilon_0 n_j + \sum_{j=1}^{D-1}t_S(f_jf_{j+1}^\dagger + f_{j+1}f_j^\dagger) + \sum_{j=1}^{D-1} U n_j n_{j+1},  
\end{equation}
where $n_j=f_j^\dagger f_j$ is the density operator of the $j$th site, and where we have assumed uniform energy bias $\epsilon_0$, hopping strength $t_S$, and nearest-neighbour repulsive interaction $U$. We characterized both leads by the same ``rounded square'' hybridization function given by 
\begin{equation}\label{eq:Jrsq}
J_\text{rsq}(\epsilon) = \frac{\Gamma}{2} \left[ \tanh\left(A\frac{\epsilon+W}{t_S}\right) - \tanh \left(A\frac{\epsilon-W}{t_S}\right) \right], 
\end{equation}
in terms of the hybridization strength $\Gamma$, the half-width $W$, and the dimensionless parameter $A$ controlling the sharpness of the band edges located at $\epsilon=\pm W$. In the limit $A\to+\infty$, $J_\text{rsq}(\epsilon)$ becomes the square hybridization function $J_\text{sq}(\epsilon) = \Gamma(\Theta(W-\epsilon)+\Theta(W+\epsilon))/2$ considered in Ref.~\cite{PhysRevX.10.031040}. By smoothing the discontinuities of the hybridization function, we can reduce the number of terms in the AAA approximation which do not contribute much to the overall physical properties of the model. A comparison with the results presented in Ref.~\cite{PhysRevX.10.031040} holds upon convergence of the simulations in the limit $A\to+\infty$.

\subsection{Single-impurity model}

To show the validity of our method, we first consider the solvable single-impurity model, which can be obtained by setting $D=1$ in the system Hamiltonian in Eq.~(\ref{eq:chainHS}). We consider a slightly challenging case, where both leads are held at zero temperature $T_L=T_R=0$ to produce discontinuous Fermi surfaces and where the parameter $A=200$ is large enough to produce sharp crossovers at the band edges, aiming to demonstrate the capability of the AAA algorithm~\cite{doi:10.1137/16M1106122,10.1063/5.0209348} in capturing such critical features. We also set different chemical potentials $\mu_L=-\mu_R=W/16$ to drive a particle current through the impurity. In Fig.~\ref{fig:impurity}(a) we show the reservoir correlation $C^{+1}(t)$ (inset) and its Fourier transform $S^{+1}(\epsilon) = 2J_\text{rsq}(\epsilon)f_L(\epsilon)$ for the left lead, alongside their approximations (dashed lines) obtained using the AAA algorithm. Both the discontinuity of $S^{+1}(\epsilon)$ at the fermi surface $\epsilon=\mu_L=W/16$ and its sharp crossover at the band edges $\epsilon=\pm W$ are reconstructed with high precision. 

For the single-impurity model considered here, in the superfermion representation~\cite{10.1063/1.3548065} the generator of the corresponding equation of motion in Eq.~(\ref{eq:ppfeom}) is quadratic in terms of the creation and annihilation operators of all system fermions and pseudofermions. This allows to solve the nonequilibrium steady-state by an exact diagonalization of the generator. More technical details can be found in Appendix~B of  Ref.~\cite{PhysRevX.10.031040}. We plot the steady-state particle current $I_{ss}$ (solid line) in Fig.~\ref{fig:impurity}(b), which is in full agreement with the analytical results (dashed line) calculated using the Landauer-B\"uttiker (LB) formula 
\begin{equation}
I_{ss}^\text{LB} = \frac{1}{2\pi}\int_{-\infty}^{+\infty} d\epsilon T(\epsilon) \big(f_L(\epsilon) - f_R(\epsilon)\big), 
\end{equation} 
where $T(\epsilon)$ is the transmission function, given by
\begin{equation}
T(\epsilon) = \frac{4\Gamma^2}{\left\lvert \det\mathbf{M}(\epsilon)\right\rvert^2}. 
\end{equation}
For the chain model considered here, the $D$-dimensional matrix $\mathbf{M}(\epsilon)$ is explicitly written as
\begin{equation}
\mathbf{M}(\epsilon) = 
\begin{bmatrix}
\epsilon-\epsilon_0-\Sigma_L(\epsilon) & t_S  & \cdots & 0 \\
t_S & \epsilon-\epsilon_0  & \cdots & 0 \\
\vdots & \vdots & \ddots & \vdots \\
0 & 0 & \epsilon-\epsilon_0 & t_S \\
0 & 0  & t_S & \epsilon-\epsilon_0-\Sigma_R(\epsilon)
\end{bmatrix}, 
\end{equation}
where 
\begin{equation}
\Sigma_L(\epsilon)=\Sigma_R(\epsilon) = -\frac{\Gamma}{\pi}\log\left\lvert\frac{\epsilon-W}{\epsilon+W}\right\rvert - i\Gamma, 
\end{equation}
are the reservoir self-energies.

Next, we focus on demonstrating the accuracy of our method. To do this, we set the parameters the same as those employed in Fig.~8 in Ref.~\cite{PhysRevX.10.031040}, i.e., $\Gamma=0.1W/8$, $T_L=1.1W$, $T_R=W/8$, $\epsilon_0=0$, and plot the steady-state particle current $I_{ss}$ as a function of the averaged chemical potential $\mu=(\mu_L+\mu_R)/2$ and the bias $V=\mu_L-\mu_R$ calculated using the LB theory in Fig.~\ref{fig:power}(a) and using the purified pseudofermion model in Fig.~\ref{fig:power}(b). Note that in the region where $VI_{ss}<0$, we set $I_{ss}=0$ by hand, following Ref.~\cite{PhysRevX.10.031040}. Our results faithfully reproduce  the prediction of the LB formula over the whole parameter region while, more impressively, avoiding the artificial oscillations visible in Fig.~8 in Ref.~\cite{PhysRevX.10.031040}, which arise from the coarse approximation of the band edges in the mesoscopic lead scheme imposed there. 

\begin{figure}
\includegraphics[clip,width=8.5cm]{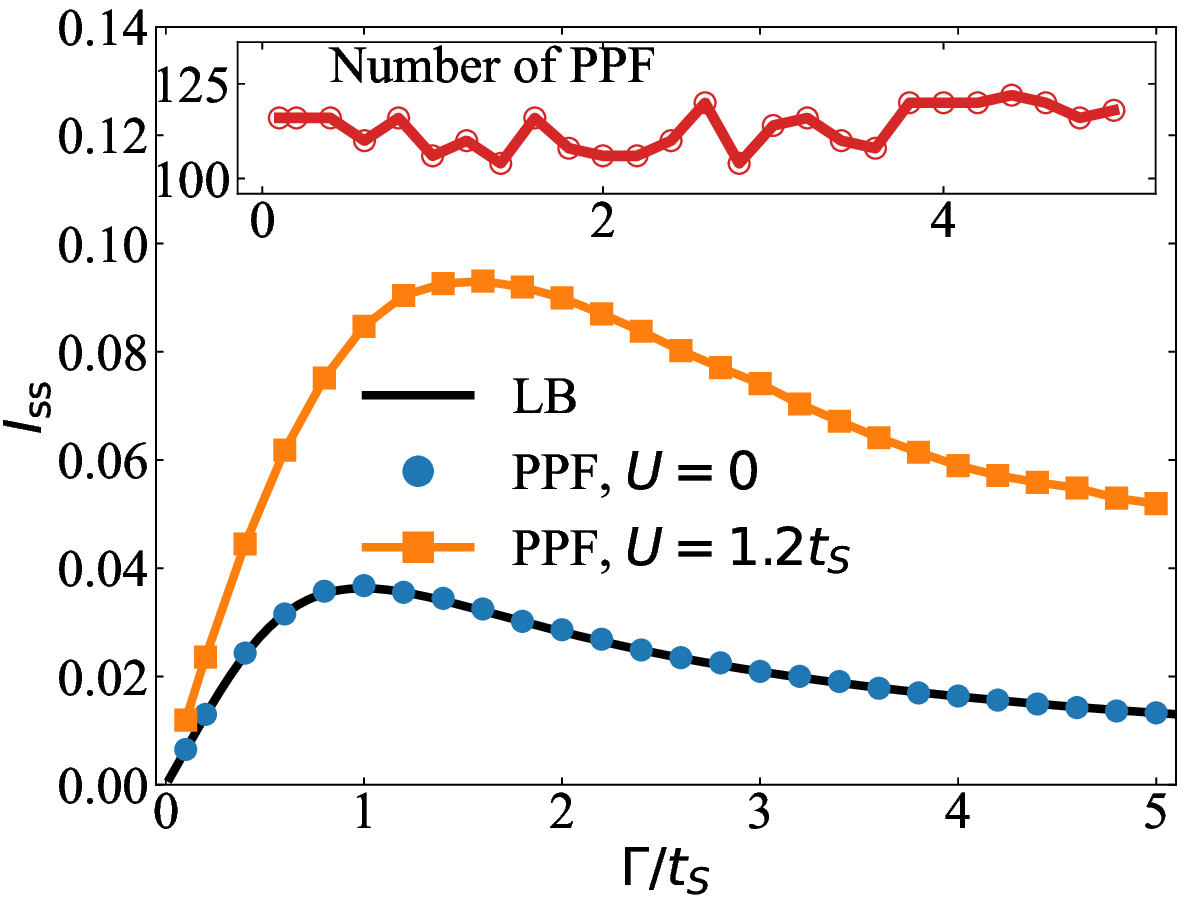}
\caption{The steady-state particle current $I_{ss}$ as a function of the hybridization strength $\Gamma$ for a three-site thermal engine in the absence $U=0$ (solid line and blue dots) and presence $U=1.2t_S$ (orange crosses) of the nearest-neighbour interaction. The relative error of the results for $U=0$ is below $2\%$, and the number of purified pseudofermions used in the simulations are shown in the inset. Other simulation parameters are $T_L=10t_S$, $T_R=t_S$, $\mu_L=-\mu_R=-t_S/2$, $W=8\Gamma$ and $A=200$. 
}\label{fig:3site}
\end{figure}

\begin{figure}
\includegraphics[clip,width=8.5cm]{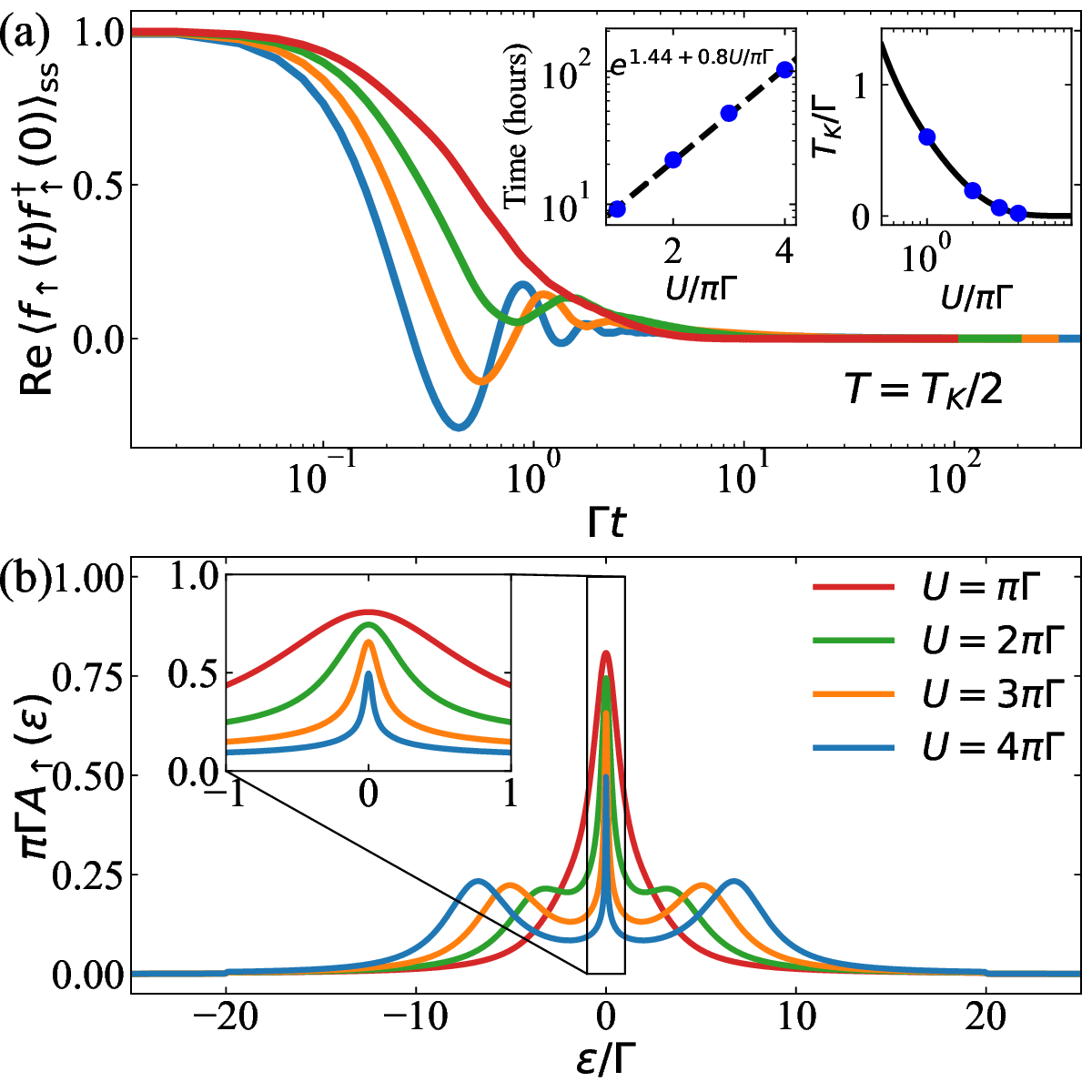}
\caption{(a) Steady-state impurity correlation function $\text{Re}\langle f_\uparrow(t)f_\uparrow^\dagger(0)\rangle_\text{ss}$ for simulation times $\Gamma T_\text{sim} = 100, 200, 300, 400$, corresponding to interaction strengths $U/\pi\Gamma= 1,2,3,4$, respectively. The reservoir temperature is set to half the Kondo temperature, $T=T_K/2$. The left inset shows the corresponding computation time (blue dots) along with an exponential fitting (dashed line). The right inset displays displays the Kondo temperature $T_K$ as a function of $U$,  with dots marking the values of $U$ used in the main figure. (b) Corresponding impurity spectral function $A_\uparrow(\epsilon)$. The inset provides a magnified view of the Kondo peaks. Other simulation parameters are $W=20\Gamma$, $A=20$ and $\mu=0$. 
}\label{fig:kondo}
\end{figure}

\subsection{Three-site thermal engine}

In this section, we demonstrate the efficiency of our method and its ability to simulate strongly-correlated systems. To this end, we consider a three-site ($D=3$) thermal engine setting with the three sites in the middle as the working medium. A temperature gradient with $T_L=10t_S$ and $T_R=t_S$ drives a current to overcome a reversed bias $\mu_L=-\mu_R=-t_S/2$ for power generation. 

In Fig.~\ref{fig:3site}, we plot the steady-state particle current $I_{ss}$ as a function of the hybridization strength $\Gamma$, both in the absence ($U=0$) and presence ($U=1.2t_S$) of the nearest-neighbour interaction. In the former case, the results (blue dots) are obtained by the exact diagonalization of the generator in Eq.~(\ref{eq:ppfeom}) and they agree with the LB formula prediction (solid line) with a relative error below $2\%$. The number of purified pseudofermions used in the numerical simulations are plotted in the inset. We observe that the number of purified pseudofermions (100--120) required to reduce the relative error below $2\%$ is almost independent of the system-lead coupling strength $\Gamma$ in a wide parameter range ($0<\Gamma\le5t_S$). 

We then use the same decompositions of the reservoir correlation functions to obtain the results (orange squares) in the presence of the nearest-neighbour interaction. In this case, in order to evolve the equation of motion in Eq.~(\ref{eq:ppfeom}) we have employed the time-dependent density renormalization group (tDMRG) algorithm augmented by the swap-gate technique to treat the long-range interactions between the purified pseudofermions and the system fermions $f_1$, $f_D$~\cite{PhysRevX.10.031040}. Since in the superfermion representation the Hilbert dimension of a purified pseudofermion is effectively $2$, the 100--120 purified pseudofermions used in our simulations correspond to 50--60 damped fermions in the mesoscopic lead scheme~\cite{PhysRevX.10.031040}. We have chosen the same parameters as those used in Fig.14(b) in Ref.~\cite{PhysRevX.10.031040}, where the data were generated with at least $100$ damped fermions to model both leads. This comparative study henceforth demonstrates a moderate improvement of the modeling efficiency of our method even for strongly correlated systems. 

\begin{figure}
\includegraphics[clip,width=8.5cm]{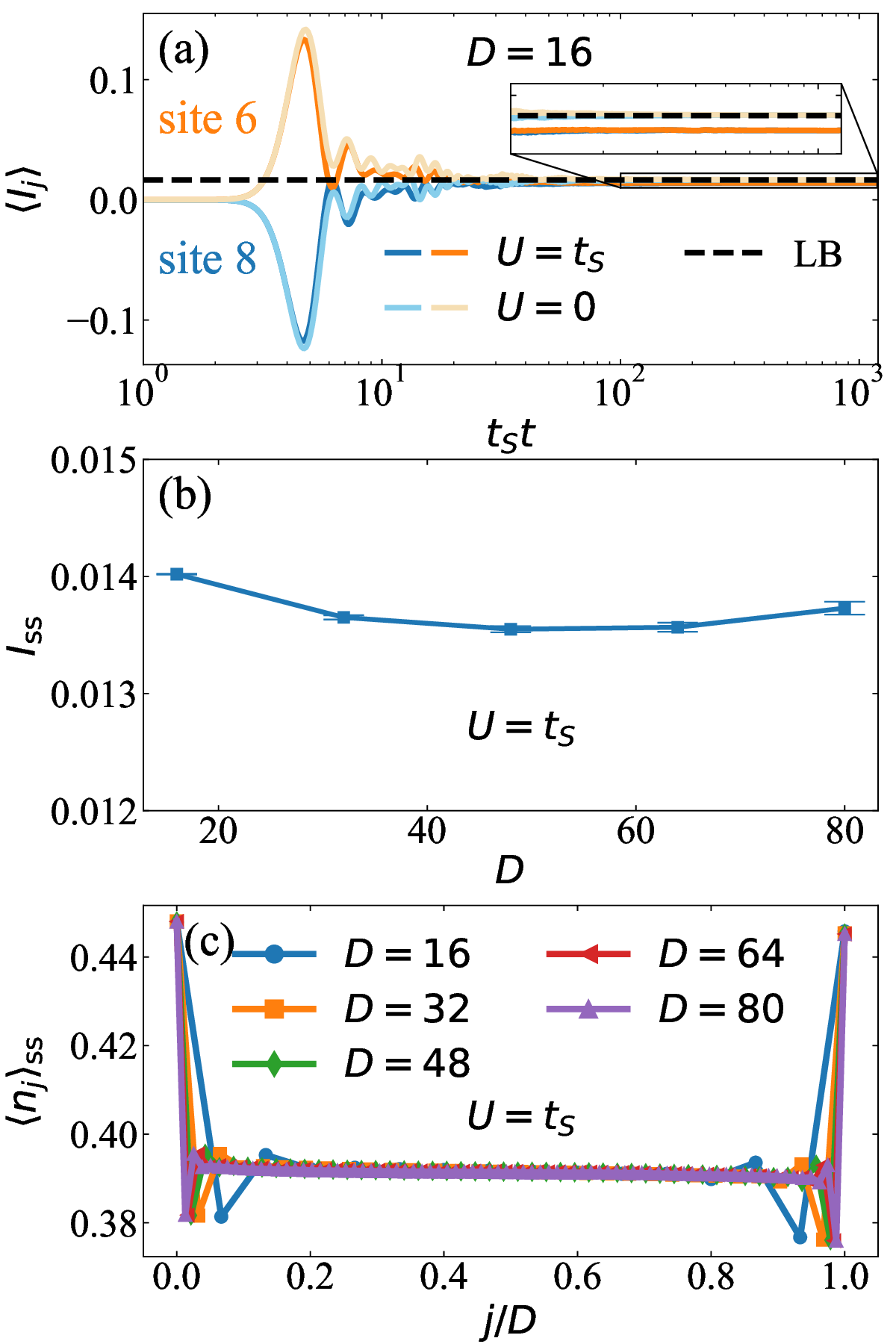}
\caption{(a) The dynamics of the particle current at the sites $j=6,8$ for a chain of length $D=16$. The inset zooms in the steady currents for better visibility. (b) Dependence of the steady-state current on the chain length $D$ for $U=t_S$. The error bars correspond to the temporal variances in the long run. (c) Distributions of the chain occupations as a function of the scaled site index $j/D$ for different chain lengths. 
}\label{fig:longchain}
\end{figure}

\subsection{Kondo resonance}
To further explore potential limitations of our approach in simulating strongly-correlated systems, we consider the equilibrium Anderson impurity model, which can be recovered by setting $D=2$ and $t_S=0$ in Eq.~(\ref{eq:chainHS}). This equivalence is made explicit through the mapping $f_1 \to f_\uparrow$, $f_2 \to f_\downarrow$, $c_{Lk} \to c_{k\uparrow}$, $c_{Rk} \to c_{k\downarrow}$, $\epsilon_{Lk} \to \epsilon_{k\uparrow}$, $g_{Lk} \to g_{k\uparrow}$, $\epsilon_{Rk} \to \epsilon_{k\downarrow}$, $g_{Rk} \to g_{k\downarrow}$, which transforms the Hamiltonian in Eq.~(\ref{eq:Hchain}) into
\begin{equation}
H = H_S + \sum_{k,s=\uparrow\downarrow} \epsilon_{k s}c_{ks}^\dagger c_{ks} + \sum_{k,s=\uparrow\downarrow} g_{ks}(f_sc_{ks}^\dagger + c_{ks}f_s^\dagger), 
\end{equation}
with the impurity Hamiltonian
\begin{equation}
H_S = \epsilon_0(f_\uparrow^\dagger f_\uparrow + f_\downarrow^\dagger f_\downarrow) + U f_\uparrow^\dagger f_\uparrow f_\downarrow^\dagger f_\downarrow.  
\end{equation}
We also assume identical chemical potentials $\mu=\mu_L=\mu_R$ and temperatures $T=T_L=T_R$, so that the impurity effectively couples to a single spinful reservoir, consistent with the standard equilibrium single-impurity Anderson model. 

In Fig.~\ref{fig:kondo}(a), we plot the steady-state impurity correlation function $\langle \{f_\uparrow(t), f_\uparrow^\dagger(0)\}\rangle_\text{ss}$ as a function of time $t$ for different repulsion strengths $U/\pi\Gamma \in\{1,2,3,4\}$. In Fig.~\ref{fig:kondo}(b), we also plot the corresponding impurity spectral function $A_s(\epsilon)$, defined as~\cite{PhysRevResearch.5.033011}
\begin{equation}
A_s(\epsilon) = \frac{1}{2\pi} \int_{-\infty}^{+\infty} dt e^{i\epsilon t}\langle \{f_s(t), f_s^\dagger(0)\}\rangle_\text{ss}, 
\end{equation} 
where $s=\,\uparrow,\downarrow$ and where $\langle \boldsymbol\cdot\rangle_\text{ss}$ represents the steady-state expectation. Here we consider the symmetric case $U = -2\epsilon_0$, so that $A_\uparrow(\epsilon) = A_\downarrow(\epsilon)$, and set the temperature to $T=T_K/2$, where the Kondo temperature $T_K$ is given by~\cite{PhysRevLett.109.266403}
\begin{equation}
T_K = \sqrt{\frac{U\Gamma}{2}}e^{-\pi U/8\Gamma + \pi\Gamma/2U}. 
\end{equation}
The right inset in Fig.~\ref{fig:kondo}(a) plots $T_K$ versus $U$, with the parameters used in the main figure indicated by blue dots. 

The plot in Fig.~\ref{fig:kondo}(b) shows that this method can reproduce a hallmark of the strong-interaction regime, i.e., the emergence of a Kondo peak at $\epsilon=0$ in the spectral function $A_\uparrow(\epsilon)$. Importantly, by choosing $T=T_K/2$, we observe that the width $W_K$ of the Kondo peak narrows when $U$ increases. Resolving this fine structure  requires to accurately compute the impurity correlation function $\langle \{f_s(t), f_s^\dagger(0)\}\rangle_\text{ss}$ up to a simulation time $T_\text{sim}$ satisfying $\pi/T_\text{sim} < W_K$. In Fig.~\ref{fig:kondo}(a), we used $\Gamma T_\text{sim} = 100, 200, 300, 400$ for $U/\pi\Gamma = 1,2,3,4$ respectively, which proves sufficient to obtain smooth Kondo peaks (see inset of Fig.~\ref{fig:kondo}(b)). The corresponding computation times are plotted in the left inset of Fig.~\ref{fig:kondo}(a), where we observe an approximate exponential scaling $T_\text{sim}\sim \exp(1.44+0.8U/\pi\Gamma)$, indicating the growing challenge of accessing larger $U$. Note that, the small kinks in $A_\uparrow(\epsilon)$ around $\epsilon = \pm20\Gamma$ stems from the band edges introduced by using $W=20\Gamma$ in Eq.~(\ref{eq:Jrsq}). 

Simulations here are performed on a commercial cluster using one AMD 7H12 CPU @ 2.6 GHz core. The number of purified pseudofermions increases with $U$ as $112,120,136,156,178$, and the bond dimension of the matrix product state was fixed at $70$.

\subsection{Transport in the chain of interacting spinless fermions}

As our last numerical demonstration, we show the capability of our method to simulate transport in extended systems. In particular, we analyze the charge transport of interacting spinless fermions in the chain model specified by Eq.~(\ref{eq:chainHS}). In Fig.~\ref{fig:longchain}(a), we show the dynamics of the particle current densities $I_j = -i(f_{j+1}f_j^\dagger - h.c.)$ at the sites $j=6$ (deep and light oranges) and $j=8$ (deep and light blues) for a chain of length $D=16$ both in the absence $U=0$ (light colors) and presence $U=t_S$ (deep colors) of the repulsive interaction, respectively. In both cases, the currents at different sites display opposite oscillations at transient times since the chosen two sites are closer to different leads. At long times, the current reaches its  steady value which agrees well with the prediction of the LB formula for $U=0$ (black dashed line). In contrast, the presence of a repulsive interaction slightly diminishes the steady-state current, see the inset. In Fig.~\ref{fig:longchain}(b), we also perform a scaling analysis of the steady-state current up to chain length $D=80$ to show the clear emergence of a ballistic regime for $D\ge32$, consistent with theoretical predictions~\cite{PhysRevX.10.031040}. The error bars in Fig.~\ref{fig:longchain}(b) represent the standard derivation $\sqrt{\int_{\Delta T}dt (I_\text{ppf}(t)-\bar{I}_\text{ppf})^2/\Delta T}$ with $\bar{I}_\text{ppf}=\int_{\Delta T} dt I_\text{ppf}(t)/\Delta T$ the time-averaged current over a time window $\Delta T=300t_S$ computed after steady-state convergence. In addition, this ballistic regime can also be seen from the distributions of site occupations in terms of the scaled coordinate $j/D$, see Fig.~\ref{fig:longchain}(c). In the bulk, site occupations converge to a slightly right-skewed distribution arising from the voltage bias $\mu_L-\mu_R>0$ when $D$ is increased.

\begin{figure}
\includegraphics[clip,width=8.5cm]{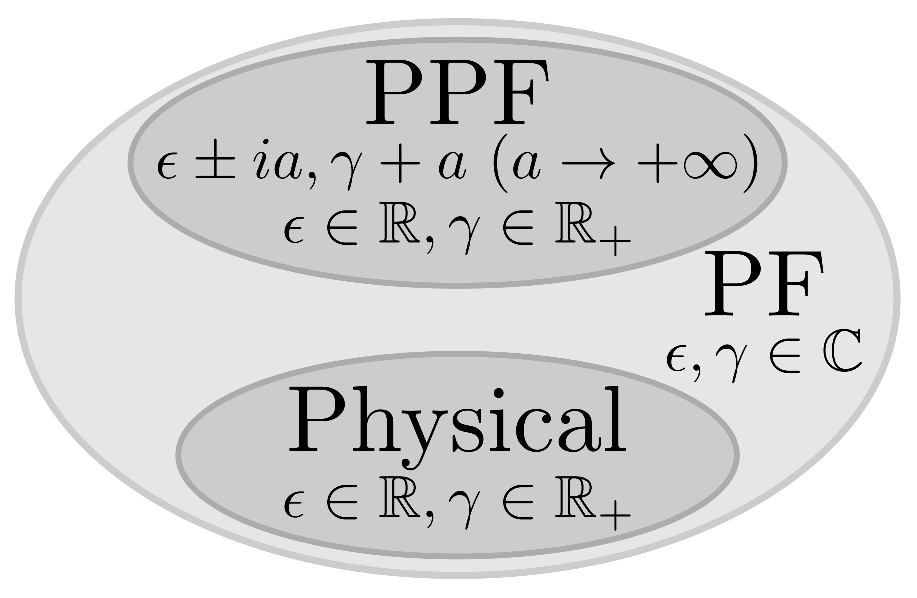}
\caption{Relations between existing methods involving ancillary damping fermions. ``Physical" refers to the auxiliary master equation approach~\cite{PhysRevLett.110.086403,PhysRevB.89.165105,PhysRevB.92.125145,PhysRevB.94.155142,PhysRevLett.121.137702,Dorda_2017,Chen_2019,PhysRevX.10.031040,PhysRevResearch.5.033011} and the mesoscopic lead scheme~\cite{PhysRevX.10.031040}, where certain constraints, e.g., real-valued fermionic energy $\epsilon\in\mathbb{R}$ and positive damping rate $\gamma\in\mathbb{R}_+$, are imposed to ensure the dynamical stability of the resulting Lindblad equations. ``PF" corresponds to the pseudofermion approach~\cite{PhysRevResearch.5.033011} which allows complex-valued energy and decay rate, i.e., $\epsilon,~\gamma\in\mathbb{C}$. ``PPF" corresponds to purified pseudofermions introduced in this paper, which can be formally understood as pseudofermions characterized by the complex energy $\epsilon\pm ia$ and the decay rate $\gamma+a$ with $\epsilon\in\mathbb{R}$ and $\gamma\in\mathbb{R}_+$, in the limit where the free parameter $a$ is sent to infinity.
}\label{fig:comparison}
\end{figure}

\section{Conclusions and outlook}\label{sec:conclusion}

In summary, we presented an efficient and accurate numerical method for the description of nonequilibrium dynamics and many-body properties in fermionic open quantum systems. 
This method relies on four types of purified pseudofermions, which are designed such that their correlation functions are compatible with powerful decomposition algorithms~\cite{10.1063/5.0209348}, as illustrated in Fig.~\ref{fig:schematics}(b). 
This compatibility allows us to take advantage of these algorithms for producing accurate approximations of the reservoir correlation functions, ultimately allowing to simulate the reduced system dynamics and nonequilibrium charge currents of the open system in a numerical exact manner.

To demonstrate the accuracy, practicality, and potential advantages  of this method with respect to other techniques, we performed a comparative numerical analysis of the charge transport of spinless fermions in a one-dimensional chain, a model previously investigated in Ref.~\cite{PhysRevX.10.031040}.
For a non-interacting chain, the results reproduced perfectly the Landauer-B\"uttiker formula, thereby confirming the validity of the method in this analytical case. We further demonstrated its accuracy also in more challenging regimes such as in the presence of discontinuities or sharp crossovers in the reservoir spectra, typically appearing at the band edges or at low temperatures. This is possible due to the compatibility  of the representation for the purified pseudofermions correlation with the AAA algorithm~\cite{doi:10.1137/16M1106122,10.1063/5.0209348} which allows to optimize the modeling efficiency over existing methods. Such compatibility is likely to be of significance in the simulation of strongly-correlated, and extended systems. To highlight this feature, we performed a scaling analysis of the steady-state current up to a chain of length $D=80$, which, to our knowledge, is inaccessible to other methods. The corresponding numerical results confirmed the ballistic characteristics of charge transport in this chain model. As a consequence, the rather optimized nature of this method holds potential significance for applications in the analysis of fermionic extended systems.

We note that the purified pseudofermion approach developed in this work, alongside its bosonic counterpart in Refs.~\cite{liang2024purifiedinputoutputpseudomodemodel,c91x-bhqw}, bears close resemblance to the dissipaton theory originally formulated in Ref.~\cite{10.1063/1.4863379} and recently extended in several directions~\cite{10.1063/5.0134700,10.1063/5.0155585,PhysRevA.110.032620,doi:10.1021/acs.jctc.5c00081}. All of these methods share a common feature: they reproduce environmental effects in an enlarged space equipped with additional ancillary degrees of freedom. The main difference lies in the physical interpretation which can be assigned to these ancillary entities. It is therefore interesting to explore the conceptual connections between these methods, in particular, whether the quasiparticles interpretation central to the dissipaton theory can be applied to the purified pseudomode/pseudofermion concepts.

We close by discussing the relation between our approach and other methods involving ancillary damped fermions, alongside their benefits and challenges encountered in practical applications. Depending on the domain of model parameters, these methods can be divided into two classes, as sketched in Fig.~\ref{fig:comparison}. 

Specifically, the ``physical'' class  refers to the auxiliary master equation approach in Refs.~\cite{PhysRevLett.110.086403,PhysRevB.89.165105,PhysRevB.92.125145,PhysRevB.94.155142,PhysRevLett.121.137702,Dorda_2017,Chen_2019,PhysRevX.10.031040,PhysRevResearch.5.033011} and the mesoscopic lead scheme in Ref.~\cite{PhysRevX.10.031040}, where the original reservoirs are mimicked by networks of independent fermions subject to Markovian damping, so that the resulting dynamics can be described by Lindblad master equations~\cite{Lindblad1976,10.1063/1.522979}, which can be efficiently integrated with advanced tensor-network algorithms. Importantly, in these models, the dynamical stability of the Lindblad equations and the existence of steady-states are ensured by imposing physical constraints such as the positivity of the Markovian damping channels ($\gamma\in\mathbb{R}_+$) and real-valued fermionic energies ($\epsilon\in\mathbb{R}$), which constitutes one of the major advantages of these methods. Nevertheless, imposing such constraints could possibly bring significant challenges in the determination of model parameters. To address this challenge, Dorda and collaborators developed a parallel tempering algorithm based on the Markov chain Monte Carlo method~\cite{Dorda_2017}, which requires possibly expensive, inefficient stochastic sampling in multi-dimensional parameter spaces. In contrast, the mesoscopic lead scheme~\cite{PhysRevX.10.031040} consists in replacing the original reservoir continuum with a discrete set of linearly or logarithmically distributed levels or their mixture. Such strategy may be expensive in computational resources and fail to capture certain features of the reservoir spectra, such as discontinuity at band edges or sharp crossovers at the Fermi surface at low temperatures as already demonstrated in Fig.~\ref{fig:impurity}, sometimes causing unwanted artifacts in the simulation results, see Fig.~8 and Fig.~9 in Ref.~\cite{PhysRevX.10.031040}.  

On the other hand, the pseudofermion approach in Ref.~\cite{PhysRevResearch.5.033011}, abbreviated as ``PF" in Fig.~\ref{fig:comparison}, lifts the constraints mentioned above by means of analytical continuation of all model parameters to the whole complex plane ($\gamma,\epsilon\in\mathbb{C}$) to enlarge model predictivity at the possible cost of numerical instability. Despite this generalization, the original pseudofermion formulation in Ref.~\cite{PhysRevResearch.5.033011} was only demonstrated for Lorentzian hybridization functions. Furthermore, since the parameters of each ancillary fermion determine the correlation in the full time-domain by construction as in Eq.(\ref{eq:Cpf}), an overhead in the number of fermionic degrees of freedom is usually required to match the output of numerical fitting tools~\cite{doi:10.1137/16M1106122,10.1063/5.0209348} that are designed for functions defined only on half of the real axis, i.e., positive or negative times.

To lift these limitations, here we defined a new purified pseudofermion model to independently model the particle/hole correlation functions in their positive- and negative-time domains. As illustrated in Fig.~\ref{fig:comparison} and detailed in Sec.~\ref{sec:ppf}, these purified versions are inherently equivalent to regular pseudofermions characterized by specific choices of the energy and decay parameters, i.e., $\epsilon\pm ia$ and $\gamma+a$ in the limit where $a$ approaches infinity. We showed that this purified representation is compatible with the AAA algorithm~\cite{doi:10.1137/16M1106122,10.1063/5.0209348} (and others) which, by allowing a direct avenue to compute the model parameters and by mitigating the numerical instabilities present in the bosonic counterpart~\cite{liang2024purifiedinputoutputpseudomodemodel,c91x-bhqw}, ultimately pushes the practicality of the pseudofermion model to more complex and realistic regimes.

\begin{acknowledgments}
M.C. acknowledges support from NSFC (Grants No. 11935012 and No. 12088101) and NSAF (Grant No. U2330401). The authors acknowledge the Beijing Super Cloud Computing Center (BSCC) for providing HPC resources that have contributed to the research results reported within this paper. 
\end{acknowledgments}

\appendix

\section{Definitions of the type I, III, IV purified pseudofermions}\label{app:ppf}

In this section, we supplement necessary details for the definitions of the type I, III, IV purified pseudofermions, corresponding to the cases (ii-iv) described in Sec.~\ref{sec:ppf}, respectively. Since most details overlap with the procedure described in Sec.~\ref{sec:ppf}, here we only summarize the differences in the derivation for each case, including the generator $L_{\text{pf},a}$ in Eq.~(\ref{eq:ac1}) and the components in its splitting $L_{\text{pf},a}=L_\text{f}+L_0+V$, the operator $S$ and, most importantly, the generating superoperator $\mathcal{L}_X$ and initial state $\rho_X(0)$ of purified pseudofermions.

\subsection{Type I: $n=1$ and negative path}
In this case, the generator $L_{\text{pf},a}$ is
\begin{equation}\label{eq:ac2}
\begin{array}{lll}
L_{\text{pf},a} 
&\displaystyle = -(\epsilon+i\gamma)dd^\dagger + (\epsilon-i(\gamma+2a))\tilde{d}\tilde{d}^\dagger \\
&\displaystyle~~~ + 2i(\gamma+a)d^\dagger\tilde{d}^\dagger + \lambda\big(sd^\dagger+ds^\dagger - \tilde{s}\tilde{d}^\dagger - \tilde{d}\tilde{s}^\dagger \big), 
\end{array} 
\end{equation}
which can be split as $L_{\text{pf},a}=L_\text{f}+L_0+V$ in terms of
\begin{equation}
\begin{array}{lll}
L_\text{f} &\displaystyle = -(\epsilon+i\gamma)dd^\dagger + \lambda (sd^\dagger + ds^\dagger), \\
L_0 &\displaystyle = (\epsilon-i(\gamma+2a))\tilde{d}\tilde{d}^\dagger + 2i(\gamma+a)d^\dagger\tilde{d}^\dagger - \lambda \tilde{s}\tilde{d}^\dagger, \\
V &\displaystyle = -\lambda \tilde{d}\tilde{s}^\dagger. 
\end{array}
\end{equation}
The operator $S$ used in the SW transformation now reads
\begin{equation}
S = -\frac{1}{2ia} \lambda\tilde{s}^\dagger\tilde{d}, 
\end{equation}
satisfying
\begin{equation}
[S, L_\text{f}+L_0] = \lambda \tilde{d}\tilde{s}^\dagger + \lambda \tilde{s}^\dagger d^\dagger + \mathcal{O}(a^{-1}), 
\end{equation}
and leading to the generator of the type I purified pseudofermion 
\begin{equation}
L_\text{I} = -(\epsilon+i\gamma)dd^\dagger + \lambda(sd^\dagger+ds^\dagger) - \lambda d^\dagger\tilde{s}^\dagger, 
\end{equation}
which in the original density matrix representation corresponds to
\begin{equation}
\begin{array}{lll}
\mathcal{L}_\text{I}[\boldsymbol{\cdot}] = - (\epsilon+i\gamma)dd^\dagger[\boldsymbol{\cdot}] + \lambda (sd^\dagger+ds^\dagger)[\boldsymbol{\cdot}] - \lambda d^\dagger[\boldsymbol{\cdot}] s.   
\end{array}
\end{equation}
Since $n=1$, the type I purified pseudofermion should take the initial state $\rho_\text{I}(0) = |1\rangle\langle1|$.

\subsection{Type III: $n=0$ and positive path}
In this case, the generator $L_{\text{pf},a}$ is
\begin{equation}\label{eq:ac3}
\begin{array}{lll}
L_{\text{pf},a} 
&\displaystyle = (\epsilon-i\gamma)d^\dagger d - (\epsilon+i(\gamma+2a))\tilde{d}^\dagger\tilde{d} \\
&\displaystyle~~~ -2i(\gamma+a)d\tilde{d} + \lambda\big(sd^\dagger+ds^\dagger - \tilde{s}\tilde{d}^\dagger - \tilde{d}\tilde{s}^\dagger \big), 
\end{array} 
\end{equation}
which can be split as $L_{\text{pf},a}=L_\text{f}+L_0+V$ in terms of
\begin{equation}
\begin{array}{lll}
L_\text{f} &\displaystyle = (\epsilon-i\gamma)d^\dagger d + \lambda (sd^\dagger + ds^\dagger), \\
L_0 &\displaystyle = -(\epsilon+i(\gamma+2a))\tilde{d}^\dagger\tilde{d} - 2i(\gamma+a)d\tilde{d} - \lambda \tilde{d}\tilde{s}^\dagger, \\
V &\displaystyle = - \lambda \tilde{s}\tilde{d}^\dagger. 
\end{array}
\end{equation}
The operator $S$ used in the SW transformation now reads
\begin{equation}
S = \frac{1}{2ia} \lambda\tilde{s}\tilde{d}^\dagger, 
\end{equation}
satisfying
\begin{equation}
[S, L_\text{f}+L_0] = \lambda \tilde{s}\tilde{d}^\dagger + \lambda \tilde{s}d + \mathcal{O}(a^{-1}), 
\end{equation}
and leading to the generator of the type II purified pseudofermion 
\begin{equation}
L_\text{III} = (\epsilon-i\gamma)d^\dagger d + \lambda(sd^\dagger+ds^\dagger) - \lambda d\tilde{s}, 
\end{equation}
which in the original density matrix representation corresponds to
\begin{equation}
\begin{array}{lll}
\mathcal{L}_\text{III}[\boldsymbol{\cdot}] = (\epsilon-i\gamma)d^\dagger d[\boldsymbol{\cdot}] + \lambda (sd^\dagger + ds^\dagger)[\boldsymbol{\cdot}] + \lambda d[\boldsymbol{\cdot}]s^\dagger.   
\end{array}
\end{equation}
Since $n=0$, the type III purified pseudofermion should take the initial state $\rho_\text{III}(0) = |0\rangle\langle0|$.

\subsection{Type IV: $n=0$ and negative path}
In this case, the generator $L_{\text{pf},a}$ is
\begin{equation}\label{eq:ac4}
\begin{array}{lll}
L_{\text{pf},a}
&\displaystyle = (\epsilon-i(\gamma+2a))d^\dagger d - (\epsilon+i\gamma)\tilde{d}^\dagger\tilde{d} \\
&\displaystyle~~~ - 2i(\gamma+a)d\tilde{d} + \lambda\big(sd^\dagger+ds^\dagger - \tilde{s}\tilde{d}^\dagger - \tilde{d}\tilde{s}^\dagger \big), 
\end{array} 
\end{equation}
which can be split as $L_{\text{pf},a}=L_\text{f}+L_0+V$ in terms of 
\begin{equation}
\begin{array}{lll}
L_\text{f} &\displaystyle = -(\epsilon+i\gamma)\tilde{d}^\dagger \tilde{d} - \lambda (\tilde{s}\tilde{d}^\dagger + \tilde{d}\tilde{s}^\dagger), \\
L_0 &\displaystyle = (\epsilon-i(\gamma+2a))d^\dagger d - 2i(\gamma+a)d\tilde{d} + \lambda ds^\dagger, \\
V &\displaystyle = \lambda sd^\dagger. 
\end{array}
\end{equation}
The operator $S$ used in the SW transformation now reads
\begin{equation}
S = - \frac{1}{2ia} \lambda sd^\dagger
\end{equation}
satisfying
\begin{equation}
[S, L_\text{f}+L_0] = - \lambda sd^\dagger + \lambda s\tilde{d} + \mathcal{O}(a^{-1}), 
\end{equation}
and leading to the generator of the type II purified pseudofermion 
\begin{equation}
L_\text{IV} = -(\epsilon+i\gamma)\tilde{d}^\dagger \tilde{d} - \lambda(\tilde{s}\tilde{d}^\dagger+\tilde{d}\tilde{s}^\dagger) + \lambda s\tilde{d}, 
\end{equation}
which in the original density matrix representation corresponds to
\begin{equation}
\begin{array}{lll}
\mathcal{L}_\text{IV}[\boldsymbol{\cdot}] = -(\epsilon+i\gamma)[\boldsymbol{\cdot}]d^\dagger d - \lambda [\boldsymbol{\cdot}](sd^\dagger + ds^\dagger) - \lambda s[\boldsymbol{\cdot}]d^\dagger.   
\end{array}
\end{equation}
Since $n=0$, the type IV purified pseudofermion should take the initial state $\rho_\text{IV}(0) = |0\rangle\langle0|$.

\bibliography{./refs}

\end{document}